\documentclass[conference, 10pt]{IEEEtran}
\IEEEoverridecommandlockouts
\usepackage{cite}
\usepackage{amsmath,amssymb,amsfonts}
\usepackage[linesnumbered, ruled]{algorithm2e}
\usepackage{graphicx}
\usepackage{textcomp}
\usepackage{xcolor}
\usepackage{amsthm}
\usepackage{mathrsfs}%
\usepackage{textcomp}%
\usepackage{manyfoot}%
\usepackage{booktabs}%
\usepackage[hidelinks]{hyperref}
\usepackage{enumerate}
\usepackage{scalerel}
\usepackage{subfigure}
\usepackage{algpseudocode}%
\usepackage{listings}%

\def\BibTeX{{\rm B\kern-.05em{\sc i\kern-.025em b}\kern-.08em
    T\kern-.1667em\lower.7ex\hbox{E}\kern-.125emX}}

\newtheorem{theorem}{Theorem}
\newtheorem{lemma}{Lemma}
\newtheorem{proposition}{Proposition}

\begin{document}

\addtolength{\textfloatsep}{-2mm}

\title{Energy-Efficient Joint Offloading and Resource Allocation for Deadline-Constrained Tasks in Multi-Access Edge Computing
\thanks{This work was supported in part by the MoE Tier-2 grant MOE-T2EP20221-0006, and in part by the National Research Foundation, Prime Minister’s Office, Singapore under its Campus for Research Excellence and Technological Enterprise (CREATE) programme.}
}

\author{\IEEEauthorblockN{Chuanchao Gao, Arvind Easwaran}
\IEEEauthorblockA{\textit{College of Computing and Data Science}}
\IEEEauthorblockA{\textit{Energy Research Institute @ NTU, Interdisciplinary Graduate Programme}}
\IEEEauthorblockA{
\textit{Nanyang Technological University, Singapore}\\
gaoc0008@e.ntu.edu.sg, arvinde@ntu.edu.sg}
}


\maketitle

\begin{abstract}
This paper addresses the deadline-constrained task offloading and resource allocation problem in multi-access edge computing. We aim to determine where each task is offloaded and processed, as well as corresponding communication and computation resource allocations, to maximize the total saved energy for IoT devices, while considering task deadline and system resource constraints. Especially, our system allows each task to be offloaded to one of its accessible access points (APs) and processed on a server that is not co-located with its offloading AP. We formulate this problem as an Integer Nonlinear Programming problem and show it is NP-Hard. To address this problem, we propose a Graph-Matching-based Approximation Algorithm ($\mathtt{GMA}$), the first approximation algorithm of its kind. $\mathtt{GMA}$ leverages linear relaxation, tripartite graph construction, and a Linear Programming rounding technique. We prove that $\mathtt{GMA}$ is a $\frac{1-\alpha}{2+\epsilon}$-approximation algorithm, where $\epsilon$ is a small positive value, and $\alpha$ ($0$$\le$$\alpha$$<$$1$) is a system parameter that ensures the resource allocated to any task by an AP or a server cannot exceed $\alpha$ times its resource capacity. Experiments show that, in practice, $\mathtt{GMA}$'s energy saving achieves $97\%$ of the optimal value on average.
\end{abstract}


\section{Introduction}\label{sec:introduction}
Recent advances in hardware, software, and communication technologies—such as ultra-reliable low-latency communication of 5G and low-power wide-area networks—have paved the way for Internet of Things (IoT) to become the next technological frontier \cite{aazam2018fog}. Emerging IoT applications, including object detection and decision-making in autonomous driving \cite{yurtsever2020survey}, are becoming increasingly computation-intensive due to the rapid evolution of Artificial Intelligence (AI) technologies. These applications pose substantial deployment challenges for battery-powered and resource-constrained IoT devices, particularly those with stringent latency requirements \cite{ramanathan2020survey}.

To address these challenges, Multi-access Edge Computing (MEC) has emerged as a promising paradigm for supporting computation-intensive and time-sensitive IoT applications. In MEC, end devices (i.e., IoT devices) can offload computation-intensive tasks to nearby Access Points (APs) via wireless networks. These tasks are subsequently forwarded to some edge server through a wired backhaul network for processing. Unlike conventional cloud computing, MEC deploys servers in close proximity to end devices, significantly reducing communication latency and enabling prompt responses to latency-sensitive tasks. While offloading tasks to MEC servers conserves energy on end devices, it also introduces communication latency and additional energy consumption associated with the offloading process. Moreover, communication and computation resources at APs and servers are limited. Therefore, an effective strategy for task mapping (to APs and servers) and resource allocation (for offloading and processing) is essential for tasks with hard deadlines.

Numerous studies have investigated deadline-constrained task offloading and resource allocation problems. Some studies assume that each task must be offloaded to a fixed AP and focus solely on mapping tasks to servers \cite{zhou2023joint, liu2023optimal, chen2020energy, dai2019learning, fan2023joint, zhao2021energy, chen2021energy, baek2020heterogeneous, yuan2020profit, fan2017deadline}. Others consider a model where each task is processed by the server co-located with the AP it is offloaded to, thereby concentrating only on task-to-AP mapping \cite{dai2018joint, dai2020edge, vu2021optimal, xu2020energy, li2019cooperative}. In the former case, bandwidth limitations at a single AP may lead to congestion during task offloading, while in the latter, the lack of flexibility in backhaul task forwarding can result in highly imbalanced server workloads. Although some studies have explored task mapping to both APs and servers \cite{gao2022deadline}, this area remains underexplored. Furthermore, existing research on deadline-constrained task offloading and resource allocation typically relies on exponential-time exact algorithms for optimal solutions \cite{vu2021optimal, fan2023joint}, or polynomial-time heuristic algorithms that lack performance guarantees \cite{zhou2023joint, liu2023optimal, chen2020energy, dai2019learning, zhao2021energy, chen2021energy, baek2020heterogeneous, yuan2020profit, fan2017deadline, dai2018joint, dai2020edge, xu2020energy, li2019cooperative}. Consequently, the domain of polynomial-time approximation algorithms—heuristics with provable performance guarantees—remains largely unexplored. Approximation algorithms not only improve computational efficiency compared to exponential-time methods but also offer performance guarantees absent in typical heuristics.

This paper addresses the deadline-constrained task offloading and resource allocation problem in MEC, aiming to maximize the total energy savings of end devices while satisfying task deadlines and system resource constraints. In our model, each task can be offloaded to one of several accessible APs (as opposed to a fixed AP). Thus, we must \textit{determine the AP to which each task is offloaded}, along with the associated communication resource allocation. Additionally, each task can be processed on a server that is not necessarily co-located with its offloading AP. Therefore, we must also \textit{determine the server on which each task is processed after offloading}, and allocate the corresponding computation resources. Moreover, to enhance the flexibility and efficiency of the offloading strategy, our model incorporates dynamic offloading power control, which allows each end device to adjust its transmission power during the offloading process. We refer to this Deadline-constrained Task offloading and Resource allocation Problem as $\mathsf{DTRP}$, and formulate it as an Integer Nonlinear Programming (INLP) problem. The Maximum Weight 3-Dimensional Matching (MW3DM) problem can be reduced to a special case of $\mathsf{DTRP}$, in which each job consume full capacity of the AP and server it is mapped to. MW3DM is NP-Hard \cite{burkard2009assignment}, implying that $\mathsf{DTRP}$ is also NP-hard. 

Furthermore, we propose the first polynomial-time approximation algorithm for $\mathsf{DTRP}$ with provable performance guarantees, termed the Graph Matching-based Approximation Algorithm ($\mathtt{GMA}$). $\mathtt{GMA}$ consists of three main steps: 1) We discretize resource allocations for tasks and formulate a Linear Programming (LP) relaxation of the original problem. 2) Based on the LP solution, we create one or more AP/server nodes for each AP/server and construct a weighted tripartite graph connecting tasks, AP nodes, and server nodes. 3) We apply an LP rounding method to derive a matching in the tripartite graph, which is mapped to a feasible solution of $\mathsf{DTRP}$. 

The main contributions of this paper are as follows:
\begin{itemize}
    \item We investigate the $\mathsf{DTRP}$ in MEC with both communication and computation contentions, aiming to maximize the total saved energy for end devices. This involves determining task mappings to both APs and servers, the resource allocations for task offloading and processing, and the power for task offloading. We formulate $\mathsf{DTRP}$ as a INLP problem and prove it is NP-Hard.
    \item Based on a novel technique to transform $\mathsf{DTRP}$ to a weighted tripartite graph matching problem, we propose the first polynomial-time approximation algorithm, $\mathtt{GMA}$, for $\mathsf{DTRP}$. We prove that $\mathtt{GMA}$ is a $\frac{1-\alpha}{2+\epsilon}$-approximation algorithm for $\mathsf{DTRP}$ (the objective obtained by $\mathtt{GMA}$ is at least $\frac{1-\alpha}{2+\epsilon}$ of the optimal objective of $\mathsf{DTRP}$), where $\epsilon$ is a small positive value, and $\alpha$ ($0$$\le$$\alpha$$<$$1$) is a system parameter indicating that any AP or server cannot allocate more than $\alpha$ times its resource capacity to any single task.
    \item We experimentally evaluate $\mathtt{GMA}$ and compare it with two existing heuristic algorithms~\cite{gao2022deadline} for $\mathsf{DTRP}$. Results show that the energy savings obtained by $\mathtt{GMA}$ is $97\%$ of the optimal value on average, while outperforming its own theoretical bound and the two heuristic algorithms by $56\%$, $22\%$ and $7\%$ on average, respectively. $\mathtt{GMA}$ outperforms the heuristic algorithms, even while those algorithms do not provide any performance guarantees.
\end{itemize}

\textbf{Paper Organization.} Section~\ref{sec:literature} surveys related work, and Section~\ref{sec:system} specifies the system model and optimization problem. Section~\ref{sec:algorithm} presents our algorithmic solution and derives its theoretical guarantee. Section~\ref{sec:experiment} presents the experiment results, and Section~\ref{sec:conclusion} concludes the paper.

\section{Literature Review}\label{sec:literature}
Due to the promising potential of MEC, the joint task offloading and resource allocation problem has received considerable attention in recent research. For a comprehensive understanding of this field, readers are referred to relevant surveys \cite{islam2021survey, saeik2021task, ramanathan2020survey, santi2021resource}. Research efforts in this domain are generally categorized into deadline-constrained problems and deadline-free problems (which typically focus on minimizing response time). In this section, we review the state-of-the-art algorithms developed for deadline-constrained task offloading and resource allocation in MEC.

The joint task offloading and resource allocation problem becomes particularly challenging when considering both bandwidth contention in wireless networks and computation resource contention at edge servers. Some studies simplify the problem by considering only computation resource contention \cite{zhou2023joint, liu2023optimal, chen2020energy, dai2018joint, dai2020edge, dai2019learning}. To tackle these simplified scenarios, researchers \cite{zhou2023joint, liu2023optimal, chen2020energy, dai2018joint} have proposed approaches that decompose the original problem into two subproblems—task mapping and resource allocation—and iteratively solve them until convergence. Others have applied reinforcement learning \cite{dai2020edge, dai2019learning}. While these methods exhibit polynomial-time complexity, they lack theoretical performance guarantees. Moreover, these studies omit bandwidth contention in MEC.

Other works explicitly address deadline-constrained task offloading and resource allocation while considering both communication and computation contentions \cite{vu2021optimal, fan2023joint, zhao2021energy, xu2020energy, li2019cooperative, chen2021energy, baek2020heterogeneous, yuan2020profit, fan2017deadline, gao2022deadline}. Some adopt exact (optimal) methods such as the branch-and-bound algorithm \cite{vu2021optimal} or Benders decomposition \cite{vu2021optimal, fan2023joint}, but their exponential time complexity limits practicality in real-world systems. To improve scalability, several studies have proposed heuristic algorithms. For instance, decomposition-based methods are used to iteratively solve task mapping and resource allocation subproblems \cite{zhao2021energy, xu2020energy, li2019cooperative}, while deep reinforcement learning approaches are adopted in \cite{chen2021energy, baek2020heterogeneous}. Other works explore meta-heuristics, including migrating birds optimization \cite{yuan2020profit} and ant colony optimization \cite{fan2017deadline}, or employ greedy strategies based on task deadlines \cite{gao2022deadline}. Although these heuristic methods offer lower computational complexity, they do not provide any theoretical performance guarantees.

Most existing studies, except \cite{gao2022deadline}, make restrictive assumptions—either fixing the offloading AP for each task \cite{zhou2023joint, liu2023optimal, chen2020energy, dai2019learning, fan2023joint, zhao2021energy, chen2021energy, baek2020heterogeneous, yuan2020profit, fan2017deadline}, or requiring task execution on the server co-located with the offloading AP \cite{dai2018joint, dai2020edge, vu2021optimal, xu2020energy, li2019cooperative}—limiting flexibility and efficiency of task execution in practical MEC deployments. We address these limitations by considering a more general model that jointly optimizes task mapping to both APs and servers, resource allocation for offloading and processing, and dynamic offloading power control under communication and computation contentions. We further propose the first polynomial-time approximation algorithm for $\mathsf{DTRP}$ with provable performance guarantees.

\section{System Model and Problem Formulation}\label{sec:system}
\subsection{MEC Architecture}\label{subsec:architecture}
\begin{figure}[t]
    \centering
    \includegraphics[page=1, width=0.8\linewidth]{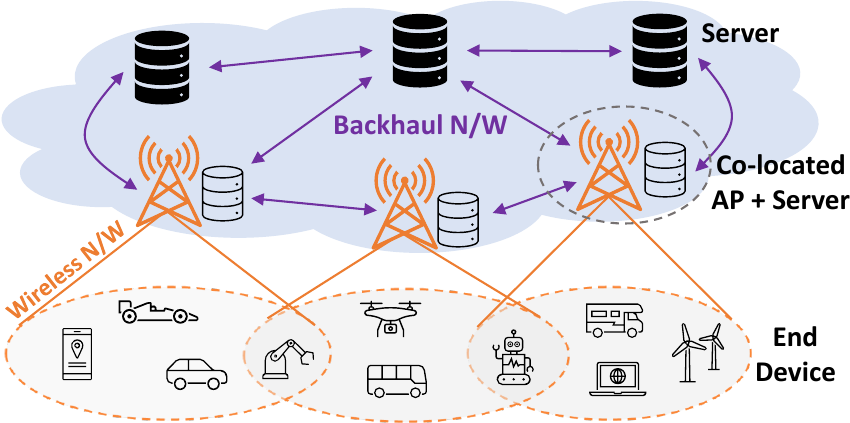}
    \caption{A typical MEC as considered in this work}
    \label{fig:system}
\end{figure}
An MEC comprises end devices, APs, and servers (Fig.~\ref{fig:system}). Tasks generated by end devices can be offloaded to nearby APs through wireless networks, and further forwarded to different servers via the wired backhaul network for processing. We denote the set of tasks as $\mathcal{I}$, where $|\mathcal{I}| = I$ ($|\cdot|$ returns the number of items in a set). Each task $i \in \mathcal{I}$ is associated with four parameters: $s_i$, $\eta_i$, and $d_i$. Here, $s_i$ represents the input size of $i$ measured in Megabits (Mb). $\eta_i$ denotes the number of CPU cycles required to process $1$ bit of input for $i$, which can be obtained by profiling task execution~\cite{alameddine2019dynamic}. $d_i$ denotes the end-to-end deadline of $i$ specified in seconds. In this paper, we consider non-splittable tasks, and each task is either fully processed locally or fully processed on some remote server. 

We denote the set of APs as $\mathcal{J}$, where $|\mathcal{J}| = J$. To better balance AP workloads, \textit{each end device can offload its tasks to one of its nearby APs}, and we denote the accessible set of APs for a task $i$ as $\mathcal{J}_i \subseteq \mathcal{J}$. \textit{In real-world systems, resource allocations are typical discrete.} We use $\bar{b}$ to denote the bandwidth unit, measured in MegaHertz (MHz), and $b_j$ to denote the bandwidth capacity for each AP $j \in \mathcal{J}$, measured in number of bandwidth units. Let $\bar{p}$ be the power unit, measured in Watts, and $p_{max}$ be the largest number of power units that end devices can use for task offloading. We denote the set of servers as $\mathcal{K}$, and denote $|\mathcal{K}| = K$. We denote the computation resource unit as $\bar{c}$, specified in CPU cycles/s, and the computation resource capacity of each server $k \in \mathcal{K}$ as $c_k$, specified in number of computation resource units. To ensure fairness in resource allocation, we introduce a user-defined system parameter $\alpha$ ($0 \le \alpha < 1$), referred to as the resource allocation bound. This parameter ensures that the resource allocated to any single task by an AP or a server cannot exceed $\alpha$ times its resource capacity. In practice, similar constraints are employed by major cloud service providers to cap the computational resources assigned to individual tasks \cite{google_cpu_limit, wang2022characterizing, aws_cpu_limit}, and by multi-antenna APs to limit the bandwidth allocated to a single device due to hardware limitations \cite{ap-wifi-limitation}. In this paper, we focus on the problem with the same $\alpha$ for both communication and computation resources. We argue that our proposed solution can also be applied to the problem where the $\alpha$ for communication resource and the $\alpha$ for computation resource are different.


\subsection{Problem Formulation}
\textit{Local Computing:} The local processing time of task $i$, $t_i^{l}$, is given by $t_i^{l} = {s_i\eta_i}/{f_i}$, where $f_i$ is the computation resource capacity available for processing task $i$ locally, specified in CPU cycles/s. In this paper, we assume tasks can meet their deadlines when they are processed locally.
The power per CPU cycle for local processing of tasks is given by $p_i^{l} = \varrho f_i^2$ \cite{han2020joint}, which is widely adopted in the literature. $\varrho$ is the energy consumption coefficient, depending on the chip architecture. Therefore, the energy consumption for local processing of $i$, $E_i^{l}$, is given as:
\begin{equation}\label{eq:model2}
    E_i^{l} = p_i^{l} \cdot s_i\eta_i = \varrho f_i^2s_i\eta_i
\end{equation}

\textit{Task Offloading:} We assume that Orthogonal Frequency Division Multiplexing (OFDM) technology is used in the wireless network. OFDM divides the network into multiple orthogonal sub-channels, which can minimize interference between tasks during offloading~\cite{li2019incentive}. Suppose task $i$ is offloaded to AP $j$. Based on Shannon's theorem \cite{shannon1984}, the task offloading rate, $r_{ij}$, can be described as:
\begin{equation}\label{eq:model3}
    r_{ij} = b_{ij} \cdot\bar{b}\cdot\log_2(1+{p_{ij}^{o}\cdot \bar{p} \cdot G_{ij}}/{\sigma^2})
\end{equation}
$b_{ij}$ is the allocated bandwidth units to task $i$, $G_{ij}$ is the channel power gain for offloading task $i$, and $\sigma$ is the average noise power. $p_{ij}^{o}$ is the power units used for offloading $i$, which cannot exceed $p_{max}$. We assume $G_{ij}$ and $\sigma$ are given for each task $i$ and each AP $j \in \mathcal{J}_i$. Thus, the offloading time of task $i$, $t_{ij}^o$, is given as 
\begin{equation}\label{eq:offload-time}
    t_{ij}^o = \frac{s_i}{r_{ij}}=\frac{s_i}{b_{ij} \cdot\bar{b}\cdot\log_2(1+{p_{ij}^{o}\cdot \bar{p} \cdot G_{ij}}/{\sigma^2})}.
\end{equation}
The consumed offloading energy, $E_{ij}^{o}$, can be computed as:
\begin{equation}\label{eq:model5}
    E_{ij}^{o} = p_{ij}^{o} \cdot \bar{p} \cdot t_{ij}^o = \frac{p_{ij}^{o} \cdot \bar{p} \cdot s_i}{b_{ij} \cdot\bar{b}\cdot\log_2(1+{p_{ij}^{o}\cdot \bar{p} \cdot G_{ij}}/{\sigma^2})}.
\end{equation}
We use $E_{ij}$ to denote the \textit{saved energy} for end devices when task $i$ is offloaded to AP $j$ given by:
\begin{equation}\label{eq:model6}
    E_{ij} = E_i^l - E_{ij}^o
\end{equation}

\textit{Task Forwarding:} Allowing tasks to be forwarded to different servers after being offloaded has advantages in balancing server workloads and mitigating wireless network coverage limitations. We consider a wired \textit{backhaul network} that connects APs and servers and has enough capacity to support data transmission with no communication contention\footnote{The bandwidth capacity of the IEEE 802.11n Wi-Fi protocol is $120$ MHz \cite{intel_wifi_protocal}, and that of a wired optical transmission system is $4.5$ THz \cite{xia2013commercial}.}. Additionally, we consider the backhaul network is enabled with Software Defined Network technology~\cite{alameddine2019dynamic}, providing monitor-based latency information among APs and servers. Thus, we assume a constant data transmission delay between a given pair of AP and server in the backhaul network~\cite{alameddine2019dynamic}, i.e., the allocated bandwidth to each task in the wired backhaul network depends on its data size. Let $\delta_{jk}$ be the delay between AP $j$ and server $k$, where $\delta_{jk} = \delta_{kj}$. When AP $j$ and server $k$ are co-located (as in Fig.~\ref{fig:system}), $\delta_{jk} = 0$.

\textit{Server Computing:} The total CPU cycles required to process task $i$ is $s_i\eta_i$, and therefore the processing time of task $i$ on server $k \in \mathcal{K}$, $t_{ik}^p$, is given by $t_{ik}^p  = s_i \cdot \eta_i/(c_{ik}\cdot \bar{c})$, where $c_{ik}$ is the allocated computation resource units to task $i$.

We use binary variables $x_{ij} \in \mathbf{x}$ and $y_{ik} \in \mathbf{y}$ to denote the offloading and processing decisions of task $i$, respectively. Specifically, $x_{ij} = 1$ if task $i$ is to be offloaded to AP $j \in \mathcal{J}_i$; otherwise, $x_{ij} = 0$. Similarly, $y_{ik} = 1$ if and only if task $i$ is to be processed on server $k \in \mathcal{K}$. Moreover, we use the integer variables $b_{ij} \in \mathbf{b}$ and $p_{ij}^o \in \mathbf{p}$ to denote the bandwidth units to be allocated to task $i$ by AP $j$ and the corresponding allocated offloading power units, respectively. We also use the integer variable $c_{ik} \in \mathbf{c}$ to denote the computation resource units to be allocated to task $i$ by server $k$. Here, the bold notation is used to denote sets of variables (i.e., $\mathbf{x}$ denotes  $\{x_{ij}| \forall i, \forall j\}$). Additionally, we use $OPT_{P}$ to denote the optimal objective value of a defined problem $P$. For convenience, we summarize major notations used in this paper in Table~\ref{table:notation}.

In this paper, we aim to find a task mapping $\langle \mathbf{x}, \mathbf{y} \rangle$ and resource allocation $\langle \mathbf{b}, \mathbf{c}, \mathbf{p}\rangle$ solution such that the total saved energy of end devices can be maximized, while satisfying the system resource and task deadline constraints. We refer to this Deadline-constrained Task offloading and Resource allocation Problem as $\mathsf{DTRP}$ and define it as follows.
\begin{subequations}
	\allowdisplaybreaks
	\begin{align}
    (\mathsf{DTRP}) \ \max \textstyle \sum\nolimits_{i \in \mathcal{I},j \in \mathcal{J}, k \in \mathcal{K}} x_{ij}y_{ik}E_{ij} \hspace{1cm} \label{eq:trp}\\
            \begin{split}
                \text{subject to:} \ \ \textstyle \sum\nolimits_{j \in \mathcal{J}_{i}}x_{ij}t_{ij}^o + \sum\nolimits_{j \in \mathcal{J}_{i}, k \in \mathcal{K}} x_{ij}y_{ik}\delta_{jk} \ \ \ \  \\
                +   \textstyle \sum\nolimits_{k \in \mathcal{K}} y_{ik}t_{ik}^p   \le d_i , \forall  i \in \mathcal{I}& \label{eq:trpa}
            \end{split}
		\\
		\textstyle \sum\nolimits_{j \in \mathcal{J}_{i}}x_{ij} \le 1 , \forall i \in \mathcal{I}& \label{eq:trpb}
		\\
		\textstyle \sum\nolimits_{j \in \mathcal{J} \setminus \mathcal{J}_{i}}x_{ij} = 0 , \forall i \in \mathcal{I}& \label{eq:trpc}
		\\
		\textstyle \sum\nolimits_{k \in \mathcal{K}}y_{ik} \le 1, \forall i \in \mathcal{I}& \label{eq:trpd}
		\\
		\textstyle \sum\nolimits_{i \in \mathcal{I}}b_{ij} \le b_j, \forall j \in \mathcal{J}& \label{eq:trpe}
		\\
		\textstyle \sum\nolimits_{i \in \mathcal{I}}c_{ik} \le c_k, \forall k \in \mathcal{K}& \label{eq:trpf}
            \\
            b_{ij} \le \alpha \cdot b_j, c_{ik} \le \alpha \cdot c_k, \ \forall i \in \mathcal{I}, \forall j \in \mathcal{J}, \forall k \in \mathcal{K}& \label{eq:trpg}
            \\
            p_{ij}^o \le p_{max}, \ \forall i \in \mathcal{I}, \forall j \in \mathcal{J}& \label{eq:trph}
		\\
		\scaleto{x_{ij} , y_{ik} \in \{0,1\}, p_{ij}^o, b_{ij}, c_{ik} \in \mathbb{Z}_{\ge 0}, \forall i \in \mathcal{I}, \forall j \in
		\mathcal{J}, \forall k \in \mathcal{K}}{10pt} & \label{eq:trpi}
	\end{align}
\end{subequations}
Constraint \eqref{eq:trpa} guarantees that each offloaded task can be completed within its deadline. Constraints \eqref{eq:trpb}$\sim$\eqref{eq:trpd} ensure that a task $i$ can only be offloaded to at most one accessible AP and processed on at most one server. Moreover, constraints \eqref{eq:trpe} and \eqref{eq:trpf} ensure that the total resource allocated to all tasks by an AP or a server cannot exceed its capacity. Finally, constraints \eqref{eq:trpg} and \eqref{eq:trph} guarantee that the resource allocation bound and offloading power bound are satisfied.
$\mathsf{DTRP}$ is an INLP problem due to the nonlinear objective function \eqref{eq:trp} and task deadline constraint \eqref{eq:trpa}. Next, we show that $\mathsf{DTRP}$ is NP-Hard in the following lemma.


\begin{table}[t]
   \caption{Notation (Key Parameters and Variables)}\label{table:notation}
   \resizebox{\columnwidth}{!}{
   \begin{tabular}{p{0.06\linewidth} p{0.83\linewidth}}
	\toprule
        \multicolumn{1}{c}{symb.} & \multicolumn{1}{c}{definition} \\
        \midrule
       $\mathcal{I}$ & task set, where $|\mathcal{I}| = I$. Each task $i \in \mathcal{I}$ is associated with input size $s_i$, CPU demand per bit data $\eta_i$, CPU cycle allocation rate for local processing $f_i$, and deadline $d_i$ \\ 
       $\mathcal{J}$ & AP set, where $j \in \mathcal{J}$ denotes an AP, and $|\mathcal{J}| = J$ \\ 
       $\mathcal{J}_i$ & set of APs to which task $i$ can be offloaded ($\mathcal{J}_i \subseteq \mathcal{J}$)   \\ 
       $\mathcal{K}$ & server set, where $k \in \mathcal{K}$ denotes a server, and $|\mathcal{K}| = K$ \\ 
       $\delta_{jk}$ & latency between AP $j$ and server $k$ in the backhaul network\\
       $b_j$ &  total number of bandwidth units ($\bar{b}$) of AP $j \in \mathcal{J}$ \\ 
       $c_k$ & total number of computation resource units ($\bar{c}$) of server $k \in \mathcal{K}$  \\ 
       $\alpha$ & resource allocation bound of all APs and servers\\ 
       $t_{ij}^o$ & offloading time of task $i$ to AP $j$\\
       $t_{ik}^p$ & processing time of task $i$ on server $k$\\
       $p_{max}$ & max offloading power units ($\bar{p}$) that can be used for any job\\ 
       $E_{i}^l$ & energy consumed for processing task $i$ locally\\ 
       $E_{ij}^o$ & energy consumed for offloading task $i$ to AP $j$\\ 
       $E_{ij}$ & energy saved by offloading task $i$ to AP $j$ for remote processing\\
       $\varphi$ & resource discretization constant; $B_m=\varphi^m$, $C_n=\varphi^n$ \\ 
       $x_{ij}$  & binary var., if task $i$ is offloaded to AP $j$; $\mathbf{x} = \{x_{ij} \mid \forall i, \forall j\}$\\ 
       $y_{ik}$  & binary var., if task $i$ is processed on server $k$; $\mathbf{y} = \{y_{ik}\mid \forall i, \forall k\}$ \\ 
       $b_{ij}$ & integer var., bandwidth units allocated to task $i$ by AP $j$; $\mathbf{b} = \{b_{ij} \mid \forall i, \forall j\}$\\ 
       $c_{ik}$ & integer var., computation resource units allocated to task $i$ by server $k$; $\mathbf{c} = \{c_{ik} \mid \forall i, \forall k\}$ \\ 
       $p_{ij}^o$ & integer var., allocated power units used to offload task $i$ to AP $j$; $\mathbf{p} = \{p_{ij}^o\mid \forall i, \forall j\}$ \\
       \bottomrule
   \end{tabular}}
\end{table}

\begin{lemma}
    $\mathsf{DTRP}$ is NP-Hard.
\end{lemma}
\begin{proof}
    The decision version of $\mathsf{DTRP}$ can be defined as follows: \textit{given an instance of $\mathsf{DTRP}$ and a target energy saving $E$, does there exist a feasible solution $\langle \mathbf{x}, \mathbf{y}, \mathbf{b}, \mathbf{c}, \mathbf{p}\rangle$ such that the total energy saving is at least $E$?} A given solution for the decision version of $\mathsf{DTRP}$ can be represented in polynomial size and verified in polynomial time with respect to the input size of $\mathsf{DTRP}$. Therefore, $\mathsf{DTRP}$ is an NP problem.

    Next, we show that $\mathsf{DTRP}$ is NP-Hard through a reduction from the MW3DM problem, known to be NP-Hard \cite{burkard2009assignment}.

    Given an instance of MW3DM: three disjoint sets $\mathcal{V}_1$, $\mathcal{V}_2$, and $\mathcal{V}_3$, a set of weighted triplets $\mathcal{T} \subseteq \mathcal{V}_1 \times \mathcal{V}_2 \times \mathcal{V}_3$, each with weight $u(v_1, v_2, v_3)$, and the goal is to find a maximum weight matching such that each element appears in at most one triplet. We construct a special case of $\mathsf{DTRP}$ as follows:
    \begin{itemize}
        \item Let each job $i$ in $\mathsf{DTRP}$ correspond to an element in $\mathcal{V}_1$, each AP $j$ in $\mathsf{DTRP}$ correspond to an element in $\mathcal{V}_2$, and each server $k$ in $\mathsf{DTRP}$ correspond to an element in $\mathcal{V}_3$.
        \item For every triplet $(i, j, k) \in \mathcal{T}$, define a feasible job mapping in $\mathsf{DTRP}$ where job $i$ is offloaded to AP $j$ and processed on server $k$.
        \item Assume that each job consumes the full capacity of the AP and server it is mapped to (i.e., resource contention constraints enforce that no AP or server can be shared across jobs).
        \item Let the energy saving of mapping job $i$ to AP $j$ and server $k$, with resource allocations $b_j$ and $c_k$, equal $u(v_1, v_2, v_3)$.
    \end{itemize}
    Then, the goal of maximizing the total energy saving in this special case of $\mathsf{DTRP}$ is equivalent to solving the MW3DM problem. Since MW3DM is NP-Hard, this special case of $\mathsf{DTRP}$ is NP-Hard. Therefore, the general $\mathsf{DTRP}$ problem, which generalizes this instance, is also NP-Hard.
\end{proof}

\section{GMA Approximation Algorithm}\label{sec:algorithm}
In MEC, tasks, APs, and servers exist as distinct and independent groups, motivating us to tackle $\mathsf{DTRP}$ using graph-matching algorithms for MW3DM. However, directly applying graph matching algorithms may lead to inefficient resource allocation in MEC, as each AP or server can only accommodate one task according to the definition of graph matching (Appendix \ref{app:matching}). To address this inefficiency, this section introduces a Graph-Matching-based Approximation Algorithm ($\mathtt{GMA}$) for $\mathsf{DTRP}$, the first polynomial-time approximation algorithm for $\mathsf{DTRP}$. 

$\mathtt{GMA}$ (Algorithm \ref{alg:approximation}) consists of three main steps. First, we perform a resource allocation discretization for reducing the number of resource allocation options and formulate a new LP problem based on $\mathsf{DTRP}$ (line $1$). Second, based on the LP solution, we construct two bipartite graphs (Appendix \ref{app:bipartite}) that establish task-to-AP mappings and task-to-server mappings, respectively. Then, we utilize sub-algorithm $\mathtt{WTGConstruct}$ to merge these two bipartite graphs into a weighted tripartite graph (Appendix \ref{app:tripartite}) whose each edge represents a task-AP-server mapping (line $2$). Third, we employ an LP rounding method to obtain a matching of the weighted tripartite graph (line $3$). This matching is then mapped to a feasible solution of $\mathsf{DTRP}$ (lines $4$--$7$). Later in this section, we will provide a detailed proof of the theoretical approximation bound for $\mathtt{GMA}$.

\subsection{Resource Allocation Discretization and LP Formulation}\label{subsec:ld}
The nonlinear objective \eqref{eq:trp} and constraint \eqref{eq:trpa} make $\mathsf{DTRP}$ challenging to solve, even when the integer variables are relaxed to continuous ones. Since both job mapping and resource allocation variables are discrete, we can enumerate all possible combinations of task mappings and resource allocations, and compute the corresponding saved energy for each. We then focus exclusively on those combinations that satisfy deadline constraints and yield positive energy savings. This approach allows us to eliminate the nonlinear task deadline constraint \eqref{eq:trpa}, while also linearizing the objective \eqref{eq:trp}, as detailed later in this subsection. 
However, the number of resource allocation options grows linearly with the value of $b_j$ and $c_k$, leading to an exponential increase in the total number of combinations as the input size of $b_j$ and $c_k$ grows. To mitigate this issue, we first apply an additional discretization to the resource allocation variables $b_{ij}$ and $c_{ik}$, thereby reducing the number of candidate options. The remainder of this subsection presents the discretization procedure and the corresponding LP formulation. Later in Theorem~\ref{theorem:gaa} (Subsection~\ref{subsec:tripartite}), we further show that the optimality gap introduced by this discretization can be effectively bounded.

We define a \emph{discretization constant}, $\varphi$, where $\varphi > 1$. For each AP $j \in \mathcal{J}$, let $\pi_j = \left\lceil \log_{\varphi}(\alpha b_j) \right\rceil$. The discretized bandwidth allocations are defined as $\{B_0, B_1, ..., B_{\pi_j}\}$, where $B_m = \left\lfloor\varphi^m \right\rfloor$ for $m = 0, ..., \pi_j-1$, and $B_{m} = \left\lfloor \alpha b_j \right\rfloor$ for $m=\pi_j$. Here, we \textit{use the logarithmic operation to ensure that $\pi_j$ (i.e., bandwidth allocation options) is polynomial in the size of input $b_j$}. Similarly, for each server $k \in \mathcal{K}$, let $\lambda_k = \left\lceil \log_{\varphi}(\alpha c_k) \right\rceil$. The discretized computation resource allocations can be defined as $\{C_0, C_1, ..., C_{\lambda_k}\}$, where $C_n = \left\lfloor \varphi^n \right\rfloor$, for $n= 0, ..., \lambda_k-1$, and $C_{n} = \left\lfloor \alpha c_k \right\rfloor$, for $n=\lambda_k$. 

\begin{algorithm}[tb]
    \caption{\scalebox{1}[1]{Graph-Matching-based Approximation (\texttt{GMA})}}\label{alg:approximation}
    \SetKwInOut{Input}{input}
    \SetKwInOut{Output}{output}
    \SetAlgoNoEnd
    \SetKwFor{ForEach}{for each}{do}{endfch}
    \SetKw{KwAnd}{and}

    Discretize task resource allocations and formulate an LP problem, $\mathsf{RDP}$, based on $\mathsf{DTRP}$. Let $\tilde{\mathbf{z}}$ be an optimal fractional solution to $\mathsf{RDP}$\;
    Construct a weighted tripartite graph $\mathcal{H}$ based on $\tilde{\mathbf{z}}$ using $\mathtt{WTGConstruct}$\;
    Formulate a relaxed maximum weighted 3-dimensional matching problem, $\mathsf{3DM}$, based on $\mathcal{H}$, and apply the $\mathtt{kDMA}$ to obtain a (integral) matching $\mathbf{M}_{\mathbf{z}}$ of $\mathcal{H}$\;

    \ForEach{$M_{\mathbf{z}}(v_i, w_{jr}, w_{ks}) = 1$}{
        $x_{ij} \leftarrow 1, y_{ik} \leftarrow 1$\;
        $b_{ij} \leftarrow b(v_i, w_{jr}, w_{ks}), c_{ik} \leftarrow c(v_i, w_{jr}, w_{ks})$\;
        Calculate $p_{ij}^o$ based on Eqs.~\eqref{eq:model5} and \eqref{eq:ld1}\;
    }
\end{algorithm}

We denote a task mapping and resource allocation combination as $\langle i,j,m,k,n \rangle$, representing that task $i$ is offloaded to AP $j$ with bandwidth allocation $B_m$ and processed on server $k$ with computation resource allocation $C_n$. 
Given a combination $\langle i,j,m,k,n \rangle$, we can compute the maximum allowable time for task offloading based on deadline constraint \eqref{eq:trpa}, i.e.,
\begin{equation}\label{eq:ld1}
    t^o_{ij} = \tau_i - \delta_{jk} - {s_i\eta_i}/{C_n}.
\end{equation}
If $t^o_{ij} > 0$, let $t^o_{ij}$ be the task offloading time. Based on $t^o_{ij}$ and bandwidth allocation $B_m$, we can compute the offloading power $p_{ij}^o$ based on \eqref{eq:offload-time}, and determine energy $E_{ij}^o$ based on Eq.~\eqref{eq:model5}. Then, we compute the saved energy associated with $\langle i,j,m,k,n \rangle$ based on Eq.~\eqref{eq:model6}, and we denote it as $E_{ijmkn}$. We claim that \textit{a combination $\langle i,j,m,k,n \rangle$ is \textbf{feasible} if the computed $t^o_{ij} > 0$, $p_{ij}^o \le p_{max}$, and $E_{ijmkn} > 0$.} According to Eq.~\eqref{eq:ld1}, the following proposition can be obtained.

\begin{proposition}\label{prop:ld1}
    A feasible combination $\langle i,j,m,k,n \rangle$ satisfies the deadline constraint \eqref{eq:trpa} of $\mathsf{DTRP}$ for task $i$.
\end{proposition}
Once resource allocations are discretized, we can enumerate all ${\langle i,j,m,k,n \rangle}$. Let $U=\max_{j \in \mathcal{J}} \pi_j$ and $V=\max_{k \in \mathcal{K}} \lambda_k$. The total number of combinations is $IJUKV$. Let $\mathcal{N}$ denote the set of all feasible $\langle i,j,m,k,n \rangle$. The time complexity to obtain the set $\mathcal{N}$ is $\mathcal{O}(IJUKV)$, which is polynomial in the input size of $\mathsf{DTRP}$. We define a new binary variable {$z_{ijmkn} \in \mathbf{z}$} for all $\langle i,j,m,k,n \rangle $, and $z_{ijmkn} = 1$ if and only if $\langle i,j,m,k,n \rangle \in \mathcal{N}$ is selected in a task offloading and resource allocation solution. We relax $z_{ijmkn}$ into a continuous variable of range $[0,1]$ and define an LP problem, denoted as $\mathsf{RDP}$, based on $\mathsf{DTRP}$. We formulate $\mathsf{RDP}$ as follows
\footnote{In equations, we use $\sum_{ i,  j, m,  k, n}$ as a shorthand notation for $\sum_{i=1}^I\sum_{j=1}^J\sum_{m=0}^{\pi_j}\sum_{k=1}^K\sum_{n=0}^{\lambda_k}$.}.
\begin{subequations}
	\allowdisplaybreaks
	\begin{align}
    (\mathsf{RDP}) \hspace{0.5cm} \max \textstyle \sum_{\langle i,j,m,k,n \rangle \in \mathcal{N}} z_{ijmkn} \cdot E_{ijmkn} \label{eq:rdp}\\
		\text{subject to: } \hspace{1cm} \textstyle \sum\nolimits_{j,m,k,n}z_{ijmkn} \le 1 ,  \ \forall i &\in \mathcal{I} \label{eq:rdpa}
            \\
		\textstyle \sum\nolimits_{ i,  m,  k,  n }z_{ijmkn}B_m \le (1-\alpha) \cdot b_j ,  \ \forall j &\in \mathcal{J} \label{eq:rdpb}
		\\
		\textstyle \sum\nolimits_{ i,  j,  m,  n}z_{ijmkn}C_n \le (1-\alpha) \cdot c_{k} ,  \ \forall k &\in \mathcal{K} \label{eq:rdpc}
		\\
		z_{ijmkn} \ge 0, \ \forall \langle i,j,m,k,n \rangle &\in \mathcal{N} \label{eq:rdpd}
	\end{align}
\end{subequations}
In $\mathsf{RDP}$, we only consider feasible combinations, so the deadline and offloading power requirements for each task are implicitly satisfied. After relaxing $\mathbf{z}$ into a continuous variable, each fractional value of $z_{ijmkn}$ represents a portion of task $i$. Eq.~\eqref{eq:rdpa} ensures that the total portions of each task $i$ do not exceed $1$. 
In the following bipartite graph construction, the AP nodes defined for each AP need at most $\alpha b_j$ \textbf{additional} bandwidth resources to accommodate their connected task nodes in a matching (as shown in the proof of Lemma \ref{lemma:bgc02}). To ensure that the final graph matching result does not violate the bandwidth constraint \eqref{eq:trpe} of $\mathsf{DTRP}$, we modify the bandwidth constraint in $\mathsf{RDP}$ as constraint \eqref{eq:rdpb} (i.e., reserve $\alpha b_j$ bandwidth for bipartite graph construction). 
The same interpretation applies to the computation resource constraint \eqref{eq:rdpc} of $\mathsf{RDP}$. The following lemma establishes a relation between the optimal solutions of $\mathsf{DTRP}$ and $\mathsf{RDP}$.

\begin{lemma}\label{lemma:ld1}
    $\frac{1-\alpha}{\varphi} {OPT_\mathsf{DTRP}} \le {OPT_\mathsf{RDP}}$.
\end{lemma}

\begin{proof}
    Assume that $\{\mathbf{x}, \mathbf{y}, \mathbf{b}, \mathbf{c}, \mathbf{p}\}$ is a feasible solution for $\mathsf{DTRP}$. Suppose an offloaded task $i$ is offloaded to AP $j$ ($x_{ij}=1$) with bandwidth allocation $b_{ij}$ and offloading power $p_{ij}^{o}$, and processed on server $k$ ($y_{ik}=1$) with computation resource allocation $c_{ik}$. Suppose the saved energy by processing $i$ on the server is $E_{ij}$. The deadline of task $i$ must be met. Besides, suppose $B_{m-1} <b_{ij} \le B_m$ and $C_{n-1} < c_{ik} \le C_{n}$ for some $m$ and $n$. As $B_{\pi_j}=\alpha b_j$ and $C_{\lambda_k}=\alpha c_k$, such $m$ and $n$ always exist. We then set $z_{ijmkn} = \frac{1-\alpha}{\varphi}$, and repeat this assignment for each offloaded task.


    Next, we show that for each $z_{ijmkn} = \frac{1-\alpha}{\varphi}$, the corresponding combination $\langle i,j,m,k,n \rangle$ is always feasible. Since $C_n \ge c_{ik}$, allocating $C_n$ computation resource will result in a shorter processing time compared with allocating $c_{ik}$ computation resource; thus, the allowable time for task offloading $t^o_{ij}$ computed in Eq. \eqref{eq:ld1} is longer. 
    Since $B_m \ge b_{ij}$, based on Eq. \eqref{eq:offload-time}, a longer offloading time and bandwidth allocation result in a smaller offloading power, i.e., the offloading power ${p}_{ijmkn}$ associated with $\langle i,j,m,k,n \rangle$ is no greater than $p_{ij}^{o}$. According to Eq.~\eqref{eq:model5}, a larger bandwidth allocation and smaller offloading power lead to a smaller offloading power consumption. This leads to a greater saved energy for $\langle i,j,m,k,n \rangle$, i.e., $E_{ijmkn} \ge E_{ij}$. Since $t^o_{ij} > 0$, ${p}_{ijmkn} \le p_{ij}^{o} \le p_{max}$, and $E_{ijmkn} \ge E_{ij} > 0$, the combination $\langle i,j,m,k,n \rangle$ is feasible. Besides, we have the following inequality.
    \[{\textstyle \sum\nolimits_{i ,j ,k}\frac{1-\alpha}{\varphi} x_{ij}y_{ik}E_{ij} \le \sum\nolimits_{\langle i,j,m,k,n \rangle \in \mathcal{N}}z_{ijmkn}E_{ijmkn}}.\]
    
    Moreover, for each offloaded task $i$, we construct only one $\langle i,j,m,k,n \rangle$ for it, which ensures the resulting $\mathbf{z}$ will satisfy constraint~\eqref{eq:rdpa} in  $\mathsf{RDP}$. Further, because $B_{m-1} < b_{ij}$, $B_{m} < b_{ij}\varphi$. Thus, for each $i \in \mathcal{I}$ and each $j \in \mathcal{J}$, we have
    \[{\textstyle \sum\nolimits_{m, k, n}z_{ijmkn}B_{m} < \frac{1-\alpha}{\varphi}x_{ij}b_{ij} \varphi = (1-\alpha)x_{ij}b_{ij}}.\]
    According to constraint~\eqref{eq:trpe} in $\mathsf{DTRP}$, $\sum_{i \in \mathcal{I}}x_{ij}b_{ij} \le b_j$. Thus, we conclude that $\sum_{i, m, k, n}z_{ijmkn}B_m \le (1-\alpha) b_j,  \ \forall j \in \mathcal{J}$, and the resulting $\mathbf{z}$ satisfies constraint~\eqref{eq:rdpb} in $\mathsf{RDP}$. Similar arguments can be used to prove that the resulting $\mathbf{z}$ also satisfies constraint~\eqref{eq:rdpc} in $\mathsf{RDP}$.
    Thus, given a feasible solution for $\mathsf{DTRP}$ with an objective value $P$, we can always construct a feasible solution for $\mathsf{RDP}$ with an objective value no less than $\frac{1-\alpha}{\varphi}P$. In particular,
    we can construct a feasible solution for $\mathsf{RDP}$ from the optimal solution of $\mathsf{DTRP}$, with an objective value $\frac{1-\alpha}{\varphi}OPT_\mathsf{DTRP}$.
\end{proof}

\begin{table}[t]
   \caption{Notation used in Section \ref{sec:algorithm}}\label{table:notation2}
   \resizebox{\columnwidth}{!}{
   \begin{tabular}{p{0.06\linewidth} p{0.83\linewidth}}
    \toprule
    \multicolumn{1}{c}{symb.} & \multicolumn{1}{c}{definition} \\
    \midrule
    $\pi_j$ & $\pi_j = \left\lceil \log_{\varphi}(\alpha b_j) \right\rceil$, number of discrete bandwidth allocation options after discretization; where $U=\max_{j\in\mathcal{J}}\pi_j$\\
    $\lambda_k$ & $\lambda_k = \left\lceil \log_{\varphi}(\alpha c_k) \right\rceil$, number of discrete computational resource allocation options after discretization; where $V=\max_{k\in\mathcal{K}}\lambda_k$\\
    $z_{ijmkn}$ & relaxed selection var. of combination $\langle i,j,m,k,n \rangle$\\
    $E_{ijmkn}$ & energy saved associated with combination $\langle i,j,m,k,n \rangle$\\
    $\tilde{z}_{ijmkn}$ & $\tilde{z}_{ijmkn} \in \mathbf{\tilde{z}}$, optimal fractional solution of LP problem $\mathsf{RDP}$\\
    $\tilde{x}_{ijm}$ & $\tilde{x}_{ijm} \in \mathbf{\tilde{x}}$, derived from Eq. \eqref{eq:bgc1} based on $\tilde{z}_{ijmkn}$\\
    $\tilde{y}_{ikn}$ & $\tilde{y}_{ikn} \in \mathbf{\tilde{y}}$, derived from Eq. \eqref{eq:bgc2} based on $\tilde{z}_{ijmkn}$\\
    $B_m$ & $m\in\{0, 1, ..., \pi_j\}$, discretized bandwidth allocation\\
    $C_n$ & $n\in\{0, 1, ..., \lambda_j\}$, discretized computation resource allocation\\
    $\mathcal{B}_{\mathbf{\tilde{x}}}$ & $\mathcal{B}_{\mathbf{\tilde{x}}} = (\mathcal{V}_{\mathbf{\tilde{x}}}, \mathcal{W}_{\mathbf{\tilde{x}}}, \mathcal{E}_{\mathbf{\tilde{x}}})$, bipartite graph constructed based on $\mathbf{\tilde{x}}$
    \\
    $\mathcal{B}_{\mathbf{\tilde{y}}}$ & $\mathcal{B}_{\mathbf{\tilde{y}}} = (\mathcal{V}_{\mathbf{\tilde{y}}}, \mathcal{W}_{\mathbf{\tilde{y}}}, \mathcal{E}_{\mathbf{\tilde{y}}})$, bipartite graph constructed based on $\mathbf{\tilde{y}}$\\
    $b(e)$ & bandwidth allocation associated with edge $e$ in a graph \\
    $c(e)$ & computational res. allocation associated with edge $e$ in a graph\\
    $\mathbf{\tilde{x}}_j$ & sorted list of $\tilde{x}_{ijm}$ defined in lines $4$--$5$ of Algorithm \ref{alg:bgc}\\
    $\tilde{x}_{j,s}$ & the $s$-th element in the sorted list $\mathbf{\tilde{x}}_j$\\
    $\mathbf{F_{\tilde{x}}}$ & fractional matching of $\mathcal{B}_{\mathbf{\tilde{x}}}$ derived from $\mathbf{\tilde{x}}$\\
    $M_{\mathbf{x}}$ & a (integral) matching of $\mathcal{B}_{\mathbf{\tilde{x}}}$\\
   \bottomrule
   \end{tabular}}
\end{table}

\subsection{Bipartite Graph Construction}\label{subsec:bigc}
$\mathsf{RDP}$ is an LP problem, efficiently solvable using algorithms such as the simplex algorithm or the ellipsoid algorithm. Leveraging an optimal solution for $\mathsf{RDP}$ enables us to establish one or more AP or server nodes for each respective AP or server. This facilitates the creation of a weighted tripartite graph, allowing multiple tasks to be potentially mapped to the same AP or server to increase resource allocation efficiency. Directly constructing the tripartite graph can be quite intricate; hence, we first utilize $\mathtt{BGConstruct}$ (Algorithm \ref{alg:bgc}) to create two bipartite graphs illustrating the task-to-AP and task-to-server mappings, respectively. In the subsequent subsection, we will introduce how to merge these two bipartite graphs into a weighted tripartite graph.

Based on an optimal solution for $\mathsf{RDP}$, denoted as $\mathbf{\tilde{z}}$, we define two variables $\tilde{x}_{ijm} \in \mathbf{\tilde{x}}$ and $\tilde{y}_{ikn} \in \mathbf{\tilde{y}}$ as follows.
\begin{align}
    \tilde{x}_{ijm} &= {\textstyle \sum\nolimits_{k, n} \tilde{z}_{ijmkn}, \forall i \in \mathcal{I}, \forall j \in \mathcal{J}, m = 0, ..., \pi_j} \label{eq:bgc1}\\
    \tilde{y}_{ikn} &= {\textstyle \sum\nolimits_{j, m} \tilde{z}_{ijmkn}, \forall i \in \mathcal{I}, \forall k \in \mathcal{K},  n = 0, ..., \lambda_k} \label{eq:bgc2}
\end{align}
As $\tilde{x}_{ijm}$ and $\tilde{y}_{ikn}$ are both in the range $[0,1]$ due to constraint~\eqref{eq:rdpa}, they represent a portion of task $i$'s mapping and resource allocation. For example, $\tilde{x}_{ijm} = \frac{1}{2}$ indicates that half of task $i$ is offloaded to AP $j$ with a discretized bandwidth allocation $B_{m}$. Based on $\mathbf{\tilde{x}}$ ($\mathbf{\tilde{y}}$), we create one or more AP (server) nodes for each AP (server) and construct bipartite graph $\mathcal{B}_{\mathbf{\tilde{x}}}$ ($\mathcal{B}_{\mathbf{\tilde{y}}}$) connecting tasks and AP (server) nodes.


Here, we provide an illustration of constructing $\mathcal{B}_{\mathbf{\tilde{x}}}$ as an example. The detailed steps for constructing $\mathbf{\mathcal{B}_{\tilde{x}}}$ and the resulting fractional matching $\mathbf{F_{\tilde{x}}}$ for $\mathbf{\mathcal{B}_{\tilde{x}}}$ are outlined in Algorithm~\ref{alg:bgc}. 
Let $\mathcal{B}_{\mathbf{\tilde{x}}} = (\mathcal{V}_{\mathbf{\tilde{x}}}, \mathcal{W}_{\mathbf{\tilde{x}}}, \mathcal{E}_{\mathbf{\tilde{x}}})$. $\mathcal{V}_{\mathbf{\tilde{x}}}$ is the set of \textit{task nodes}, and $\mathcal{W}_{\mathbf{\tilde{x}}}$ is set of \textit{AP nodes} (lines $1$--$2$). By summing up task fractions mapped to AP $j$, we determine the number of AP nodes that are defined for AP $j$, which is denoted as $n_j$ and given by
\begin{equation}\label{eq:bgc01}
    n_j \triangleq \left\lceil {\textstyle \sum\nolimits_{i, m} \tilde{x}_{ijm}} \right\rceil, \forall j \in \mathcal{J}.
\end{equation}
Besides, for each edge $e=(v_i,w_{jr}) \in \mathcal{E}_{\mathbf{\tilde{x}}}$, let $b(e) \in \mathbf{b}$ denotes its associated bandwidth allocation. For each AP $j$, we first sort all positive $\tilde{x}_{ijm}$, for $i \in \mathcal{I}, m \in \{0,1, ...,\pi_j\}$, in non-increasing order of $m$, where a larger $m$ represents a larger bandwidth allocation $B_m$. Let $\mathbf{\tilde{x}}_j$ denote this sorted list, and $\tilde{x}_{j,s}$ denote the $s$-th element in $\mathbf{\tilde{x}}_j$ (lines $4$--$5$). 

If $ n_j = 1$, we create a single AP node $w_{j1}$ for AP $j$ and establish connections between this AP node and all the tasks with positive $\tilde{x}_{ijm}$ (lines $6$--$9$). 

\begin{figure}[t]
    \centering
    \includegraphics[page=3, width=\columnwidth]{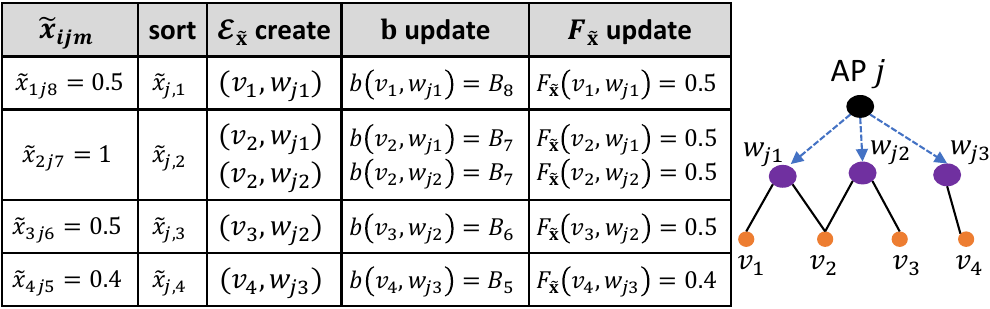}
    \caption{Example of edge creation for AP $j$ when $n_j > 1$: Given that $\sum\tilde{x}_{ijm} = 2.4$, we create three AP nodes for AP $j$ (where $i$ denotes the task node index and $m$ indicates the bandwidth allocation level $B_m$). The values of $\tilde{x}_{ijm}$ are sorted in descending order of $m$, and we denote the $s$-th element in the sorted list as $\tilde{x}_{j,s}$. Edges between task nodes and AP nodes are then constructed based on this sorted list, with AP nodes ordered as $w_{j1}, w_{j2}, w_{j3}$. The first element, $\tilde{x}_{j,1}$, corresponds to $\tilde{x}_{1j8}$. Therefore, we create an edge $(v_1, w_{j1})$ between task node $v_1$ and AP node $w_{j1}$, with bandwidth allocation $b(v_1, w_{j1}) = B_8$ and edge fraction $F_{\tilde{\mathbf{x}}}(v_1, w_{j1}) = \tilde{x}_{1j8} = 0.5$. The second element, $\tilde{x}_{j,2}$, corresponds to $\tilde{x}_{2j7}$. Since $\tilde{x}_{1j8} + \tilde{x}_{j,2} > 1$, define two edges based on $\tilde{x}_{j,2}$, $(v_2, w_{j1})$ and $(v_2, w_{j2})$. Specifically, we ensure that the total fraction assigned to each AP node does not exceed $1$. For example, $F_{\tilde{\mathbf{x}}}(v_1, w_{j1}) + F_{\tilde{\mathbf{x}}}(v_2, w_{j1}) = 1$, and the remaining portion is assigned as $F_{\tilde{\mathbf{x}}}(v_2, w_{j2}) = \tilde{x}_{1j8} + \tilde{x}_{j,2} - 1 = 0.5$. The edge construction process continues similarly for the remaining elements $\tilde{x}_{j,s}$ in the sorted list.
    }  
    \label{fig:gma2}
\end{figure}

If $n_j >1$ (e.g., Fig.~\ref{fig:gma2}), we create $n_j$ AP nodes for AP $j$. We then traverse the sorted list $\mathbf{\tilde{x}}_j$ from left to right, and link task nodes to nodes of AP $j$ such that the sum of fractions ($F_{\mathbf{\tilde{x}}}(e)$) assigned to edges incident on each AP node is exactly $1$ (line $13$). \textit{This ensures that each AP node can accommodate exactly one task node and prevents resource over-provisioning.} For $s$-element in $\mathbf{\tilde{x}}_j$, $\tilde{x}_{j,s}$, we first compute $\sum_{l = 1}^{s-1}{\tilde{x}}_{j,l}$ to determine which AP node a task node should link to (line $11$). Each $\tilde{x}_{j,s}$ can create one or two edges, depending on the value of $\sum_{l = 1}^{s}{\tilde{x}}_{j,l}$ (lines $12$--$18$).

The bipartite graph $\mathcal{B}_{\mathbf{\tilde{y}}} = (\mathcal{V}_{\mathbf{\tilde{y}}}, \mathcal{W}_{\mathbf{\tilde{y}}}, \mathcal{E}_{\mathbf{\tilde{y}}})$, $\mathbf{c}$, and a fractional matching $F_{\mathbf{\tilde{y}}}$ of $\mathcal{B}_{\mathbf{\tilde{y}}}$ can be obtained using similar steps. Here, $\mathcal{V}_{\mathbf{\tilde{y}}} = \{v_i: i = 1, ..., I\}$ is the set of task nodes, and $\mathcal{W}_{\mathbf{\tilde{y}}} = \{w_{ks}: k = 1, ..., K, s = 1, ..., n_k\}$ is the set of server nodes, where $n_k = \left\lceil \sum_{i ,n} \tilde{y}_{ikn} \right\rceil$. Besides, $\mathbf{c}$ is the set of computation resource allocation associated with each edge in $\mathcal{E}_{\mathbf{\tilde{y}}}$. 

\textbf{Time Complexity.} For each AP $j$, there are at most $I$ tasks and $U$ discrete bandwidth allocation options. Therefore, $\mathbf{\tilde{x}}_{j}$ contains at most $IU$ elements. Sorting $\mathbf{\tilde{x}}_{j}$ has a time complexity of $\mathcal{O}(IU\log(IU))$, and at most two edges are created for each $\tilde{x}_{ijm} > 0$ (lines $12$--$18$). There are at most $J$ APs. Therefore, the time complexity of $\mathtt{BGConstruct}$ for constructing $\mathcal{B}_{\mathbf{\tilde{x}}}$ is $\mathcal{O}(IJU\log(IU))$. Similarly, for each server $k$, there are at most $I$ tasks and $V$ discrete computational resource allocation options. The time complexity of $\mathtt{BGConstruct}$ for constructing $\mathcal{B}_{\mathbf{\tilde{y}}}$ is $\mathcal{O}(IKV\log(IV))$.

In the following lemma, we show that the total resource allocations corresponding to any matching of $\mathcal{B}_{\mathbf{\tilde{x}}}$ (or $\mathcal{B}_{\mathbf{\tilde{y}}}$) satisfies the resource constraints~\eqref{eq:trpe} (or \eqref{eq:trpf}) of $\mathsf{DTRP}$. This property will then be used to show that any matching of the weighted tripartite graph constructed in Subsection~\ref{subsec:tripartite} will also satisfy the resource constraints of $\mathsf{DTRP}$.


\begin{algorithm}[tb]
\caption{$\mathtt{BGConstruct}$}\label{alg:bgc}
\SetAlgoNoEnd
\SetKwInOut{Input}{input}
\SetKwInOut{Output}{output}
\SetKwFor{ForEach}{for each}{do}{endfch}
\SetKw{KwAnd}{and}
\SetKwFunction{FEdgeAssign}{\textbf{Assign}}
\SetKwProg{Fn}{Function}{:}{}
\SetNoFillComment

\Input{$\mathbf{\tilde{x}}$ (\small{\textit{obtained from Eq.\eqref{eq:bgc1}}})}
\Output{$\mathcal{B}_{\mathbf{\tilde{x}}} = (\mathcal{V}_{\mathbf{\tilde{x}}}, \mathcal{W}_{\mathbf{\tilde{x}}}, \mathcal{E}_{\mathbf{\tilde{x}}})$, $\mathbf{b}$}
\BlankLine
Initialize $\mathcal{B}_{\mathbf{\tilde{x}}} = (\mathcal{V}_{\mathbf{\tilde{x}}}, \mathcal{W}_{\mathbf{\tilde{x}}}, \mathcal{E}_{\mathbf{\tilde{x}}})$ and $\mathbf{b}$\;
$\mathcal{W}_{\mathbf{\tilde{x}}}\leftarrow \{w_{jr} \mid j = 1, ..., J, r = 1, ..., n_j\}$, $\mathcal{V}_{\mathbf{\tilde{x}}} \leftarrow \{v_i\mid i = 1, ...,  I\}$, $\mathcal{E}_{\mathbf{\tilde{x}}} \leftarrow \emptyset$\;
\ForEach{$j \in \mathcal{J}$}{
    $\mathbf{\tilde{x}}_j \leftarrow \{\tilde{x}_{ijm} \in \mathbf{\tilde{x}} \mid i \in \mathcal{I}, m = 0, ..., \pi_j, \tilde{x}_{ijm} > 0\}$\;
    sort $\mathbf{\tilde{x}}_j$ in non-increasing order of $m$ values (\small{\textit{ties broken arbitrary; $\tilde{x}_{j,s}$ denotes the $s$-th element in $\mathbf{\tilde{x}}_j$}})\;
    \textbf{if} $n_j == 1$, \For{$s \leftarrow 1$ \KwTo $|\mathbf{\tilde{x}}_j|$}{
            Suppose $\tilde{x}_{j,s}$ corresponds to $\tilde{x}_{ijm}$\;
            \FEdgeAssign{$(v_i, w_{j1}), B_m, \tilde{x}_{ijm}$}\;
            \textbf{return}\;
        }
    
    \tcc{\textcolor{darkgray}{for the case of $n_j > 1$}}
    \For{$s \leftarrow 1$ \KwTo $|\mathbf{\tilde{x}}_j|$}{
        Suppose $r-1\le \sum_{l = 1}^{s-1}{\tilde{x}}_{j,l} < r$ for integer $r$\;
        \If{$\sum_{l = 1}^{s}{\tilde{x}}_{j,l} \le r$}
        {
            \tcc{\textcolor{darkgray}{\scalebox{.75}[1.0]{\small{create one edge for $\tilde{x}_{j,s}$ linked to $w_{jr}$}}}}
            Suppose $\tilde{x}_{j,s}$ corresponds to $\tilde{x}_{ijm}$\;
            \FEdgeAssign{$(v_i, w_{jr}), B_m, \tilde{x}_{ijm}$}\;
        }
        \Else{
            \tcc{\textcolor{darkgray}{\footnotesize{create two edges for $\tilde{x}_{j,s}$, one to $w_{jr}$, one to $w_{j,r+1}$}}}
            Suppose $\tilde{x}_{j,s}$ corresponds to $\tilde{x}_{ijm}$\;
            \FEdgeAssign{$(v_i, w_{jr}), B_m, r - \sum_{l = 1}^{s-1}{\tilde{x}}_{j,l}$}\;
            \FEdgeAssign{$(v_i, w_{j,r+1}), B_m, \sum_{l = 1}^{s}{\tilde{x}}_{j,l} - r$}
        }
    }
}
\BlankLine
\Fn{\FEdgeAssign{$e, b, x$}}{
    \lIf{$ e \notin \mathcal{E}_{\mathbf{\tilde{x}}}$}{
        $\mathcal{E}_{\mathbf{\tilde{x}}} \leftarrow \mathcal{E}_{\mathbf{\tilde{x}}} \cup \{e\}, {b}(e) \leftarrow b$
    }
    \scalebox{.9}[0.9]{{\textbf{if} $ e \notin \mathcal{E}_{\mathbf{\tilde{x}}}, \scaleto{F}{6pt}_{\mathbf{\tilde{x}}}(e) \leftarrow x$; 
     \textbf{else} $\scaleto{F}{6pt}_{\mathbf{\tilde{x}}}(e) \leftarrow \scaleto{\mathcal{F}}{6pt}_{\mathbf{\tilde{x}}}(e) + x$}}; 
}
\end{algorithm}

\begin{lemma}\label{lemma:bgc02}
    Suppose $\mathbf{M}_{\mathbf{x}}$ is any (integral) matching of $\mathcal{B}_{\mathbf{\tilde{x}}}$, and $\mathbf{M}_{\mathbf{y}}$ is any matching of $\mathcal{B}_{\mathbf{\tilde{y}}}$. Then, the total allocated bandwidth or computation resource by an AP or a server does not exceed its resource capacity, i.e.,
    \begin{align*}
        &{\textstyle \sum_{i=1}^I\sum_{r=1}^{n_j} M_{\mathbf{x}}(v_i, w_{jr}) {b}(v_i, w_{jr}) \le b_j, \forall j \in \mathcal{J}};\\
        &{\textstyle \sum_{i=1}^I\sum_{s=1}^{n_k} M_{\mathbf{y}}(v_i, w_{ks}){c}(v_i, w_{ks}) \le c_k, \forall k \in \mathcal{K}}.
    \end{align*}
\end{lemma}

\begin{proof}
    We apply $\mathtt{BGConstruct}$ to construct two bipartite graphs ($\mathcal{B}_{\mathbf{\tilde{x}}}$ and $\mathcal{B}_{\mathbf{\tilde{y}}}$) based on the LP solution of $\mathsf{RDP}$. This construction method is inspired by the approach of Shmoys and Tardos \cite{shmoys1993approximation}, who used it to build a bipartite graph from the relaxed LP solution of a one-dimensional GAP. They proved (Theorem 2.1 in \cite{shmoys1993approximation}) that the total resource usage of an integral matching (e.g., $\mathbf{M}_{\mathbf{x}}$) in the constructed bipartite graph (e.g., $\mathcal{B}_{\mathbf{\tilde{x}}}$) does not exceed the resource capacity specified in the LP (e.g., $(1-\alpha) b_j$ in constraint \eqref{eq:rdpb}) plus the maximum allocation allowed for a single task (e.g., $\alpha b_j$ in constraint \eqref{eq:trpg}). Although our problem allows different resource allocations per task, the result of Shmoys and Tardos remains applicable. We omit the formal proof here for brevity. Consequently, the total bandwidth demand of $\mathbf{M}_{\mathbf{x}}$ on any AP $j$ is at most $(1-\alpha) b_j + \alpha b_j = b_j$. Similarly, the total computation demand in $\mathcal{B}_{\mathbf{\tilde{y}}}$ on any server $k$ is at most $c_k$.
\end{proof}

\subsection{Weighted Tripartite Graph Construction}\label{subsec:tripartite}

\begin{algorithm}[t]
    \caption{$\mathtt{WTGConsturct}$}\label{alg:tgc}
    \SetAlgoNoEnd
    \SetKwInOut{Input}{input}
    \SetKwInOut{Output}{output}
    \SetKwFor{ForEach}{for each}{do}{endfch}
    \SetKw{KwAnd}{and}

    \Input{$\tilde{\mathbf{z}}$ (\textit{optimal solution for $\mathsf{RDP}$})}
    \Output{$\mathcal{H} = (\mathcal{V}_1, \mathcal{V}_2, \mathcal{V}_3, \mathcal{E}), \mathbf{b}, \mathbf{c}, \mathbf{u}$}
    \BlankLine

    Initialize $\mathcal{H} = (\mathcal{V}_1, \mathcal{V}_2, \mathcal{V}_3, \mathcal{E})$, $\mathbf{b}$, $\mathbf{c}$ and $\mathbf{u}$ \;
    Calculate $\mathbf{\tilde{x}}$ and $\mathbf{\tilde{y}}$ based on Eqs.~\eqref{eq:bgc1} and \eqref{eq:bgc2}\;
    $\mathcal{B}_{\mathbf{\tilde{x}}}= (\mathcal{V}_{\mathbf{\tilde{x}}}, \mathcal{W}_{\mathbf{\tilde{x}}}, \mathcal{E}_{\mathbf{\tilde{x}}}) \leftarrow \mathtt{BGConstruct}(\mathbf{\tilde{x}})$\;
    $\mathcal{B}_{\mathbf{\tilde{y}}}= (\mathcal{V}_{\mathbf{\tilde{y}}}, \mathcal{W}_{\mathbf{\tilde{y}}}, \mathcal{E}_{\mathbf{\tilde{y}}})\leftarrow \mathtt{BGConstruct}(\mathbf{\tilde{y}})$\;
    $\mathcal{V}_1 \leftarrow \mathcal{V}_{\mathbf{\tilde{x}}}, \mathcal{V}_2 \leftarrow \mathcal{W}_{\mathbf{\tilde{x}}}, \mathcal{V}_3 \leftarrow  \mathcal{W}_{\mathbf{\tilde{y}}}, \mathcal{E} \leftarrow \emptyset$\;

    Sort $\tilde{\mathbf{z}}$ in non-increasing order of $E_{ijmkn}$ values (\textit{saved energy of combination $\langle i,j,m,k,n \rangle$})\;
    \ForAll{$\tilde{z}_{ijmkn} \in \tilde{\mathbf{z}}, \tilde{z}_{ijmkn} > 0$}{
        Let $\mathcal{E}_{ijm}$ be the set containing the one or two edges in $\mathcal{E}_{\mathbf{\tilde{x}}}$ created based on $\tilde{x}_{ijm}$, i.e., $\tilde{x}_{j,s}$ (created in lines $12$--$18$ of $\mathtt{BGConstruct}$)\;
        Let $\mathcal{E}_{ikn}$ be the set containing the one or two edges in $\mathcal{E}_{\mathbf{\tilde{y}}}$ created based on $\tilde{y}_{ikn}$, i.e., $\tilde{y}_{k,s}$ (created in lines $12$--$18$ of $\mathtt{BGConstruct}$)\;
        \ForEach{$(v_i, w_{jr}) \in \mathcal{E}_{ijm}, (v_i, w_{ks}) \in \mathcal{E}_{ikn}$}{
            Define edge $e = (v_{i}, w_{jr}, w_{ks})$\;
            \If{$ e \notin \mathcal{E}$}{
                    $\mathcal{E} \leftarrow \mathcal{E} \cup \{e\}$,
                    $ u(e) \leftarrow E_{ijmkn}, b(e) \leftarrow {b}(v_{i}, w_{jr}), c(e) \leftarrow {c}(v_{i}, w_{ks})$\;
                }
        }
    }
\end{algorithm}

The constructed bipartite graphs $\mathcal{B}_{\mathbf{\tilde{x}}}$ and $\mathcal{B}_{\mathbf{\tilde{y}}}$ have the same set of task nodes, which have a one-to-one correspondence with all the tasks. Thus, we utilize $\mathtt{WTGConstruct}$ (Algorithm~\ref{alg:tgc}) to define a weighted tripartite graph $\mathcal{H} = (\mathcal{V}_1 \cup \mathcal{V}_2 \cup \mathcal{V}_3, \mathcal{E})$ by merging $\mathcal{B}_{\mathbf{\tilde{x}}}$ and $\mathcal{B}_{\mathbf{\tilde{y}}}$. For ease of presentation, we also use $\mathbf{b}$ and $\mathbf{c}$ to denote the communication and computation resource allocations associated with each edge in $\mathcal{E}$, and use $\mathbf{u}$ to denote the weight assigned to each edge in $\mathcal{E}$.
Let $\mathcal{V}_1$ contain all the task nodes, $\mathcal{V}_2$ contain all the AP nodes, and $\mathcal{V}_3$ contain all the server nodes (line $5$). We first sort $\tilde{\mathbf{z}}$ in non-increasing order of $E_{ijmkn}$ values (ties broken arbitrarily).
Each $\tilde{z}_{ijmkn} > 0$ will result in a $\tilde{x}_{ijm} > 0$ based on Eq.~\eqref{eq:bgc1}, which can define to at most two edges in $\mathcal{B}_{\mathbf{\tilde{x}}}$ based on lines $12$--$18$ of $\mathtt{BGConstruct}$. Similarly, each $\tilde{z}_{ijmkn} > 0$ will result in a $\tilde{y}_{ikn} > 0$ based on Eq.~\eqref{eq:bgc2}, which can define to at most two edges in $\mathcal{B}_{\mathbf{\tilde{y}}}$. Therefore, for each $\tilde{z}_{ijmkn} > 0$, we can determine the corresponding edge sets $\mathcal{E}_{ijm}$ and $\mathcal{E}_{ikn}$ (lines $8$--$9$). Then, for each possible merged edge $e$, if it has not been added to set $\mathcal{E}$, we add it to set $\mathcal{E}$, and assign its weight $u(e)$, bandwidth allocation $b(e)$, and computation resource allocation $c(e)$ (lines $10$--$13$).

\textbf{Time Complexity.} The time complexity of $\mathtt{BGConstruct}$ for constructing $\mathcal{B}_{\mathbf{\tilde{x}}}$ and $\mathcal{B}_{\mathbf{\tilde{y}}}$ are $\mathcal{O}(IJU\log(IU))$ and $\mathcal{O}(IKV\log(IV))$, respectively. The number of positive $\tilde{z}_{ijmkn}$ is at most $IJUKV$, and the time complexity for sorting them is $\mathcal{O}(IJUKV\log(IJUKV))$. Each positive $\tilde{z}_{ijmkn}$ can define at most $4$ edges. Therefore, the time complexity of $\mathtt{WTGConstruct}$ is $\mathcal{O}(IJUKV\log(IJUKV))$.

The following two lemmas summarize some properties of the constructed weighted tripartite graph.
feasible combinations $\langle i,j,m,k,n \rangle$ can have $\tilde{z}_{ijmkn} > 0$, and Lemma~\ref{lemma:tgc2} holds because of Lemma~\ref{lemma:bgc02}.

\begin{lemma}\label{lemma:tgc1}
    $\forall (v_i, w_{jr}, w_{ks}) \in \mathcal{E}$, the deadline of task $i$ can be met with the task-AP-server combination $(i,j,k)$ and resource allocations $(b(v_i, w_{jr}, w_{ks}), c(v_i, w_{jr}, w_{ks}))$.
\end{lemma}
\begin{proof}
    Based on line $7$ of $\mathtt{WTGConstruct}$, we only construct an edge $\forall (v_i, w_{jr}, w_{ks}) \in \mathcal{E}$ when the corresponding $\tilde{z}_{ijmkn} \ge 0$. We only define variable $\tilde{z}_{ijmkn}$ when the combination $\langle i,j,m,k,n \rangle$ is feasible, which meets the task deadline requirement (Proposition \ref{prop:ld1}). Based on line $13$ of $\mathtt{WTGConstruct}$, we set $b(v_i, w_{jr}, w_{ks}) = b(v_i, w_{jr})$, where $b(v_i, w_{jr}) \ge B_m$ based on line $20$ of $\mathtt{BGConstruct}$. Similarly, we have $c(v_i, w_{jr}, w_{ks}) \ge C_n$. Given the task mapping to AP $j$ and server $k$, since the deadline of task $i$ can be met with resource allocation $B_m$ and $C_n$, the deadline of task $i$ can also be met with resource allocation ${b}(v_{i}, w_{jr})$ and ${c}(v_{i}, w_{ks})$ as the same or more resource are allocated.
\end{proof}

\begin{figure*}[tb!]
    \centering
    \includegraphics[page=1, width=0.8\textwidth]{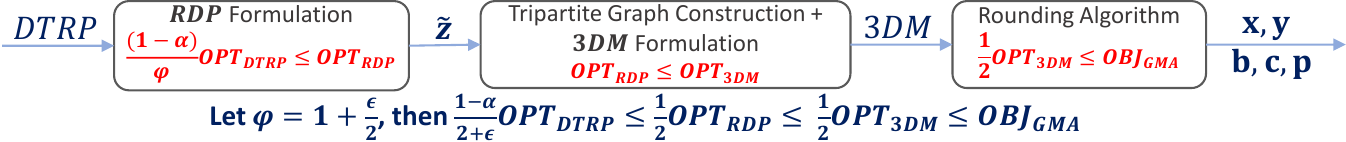}
    \caption{$\mathtt{GMA}$ flow and its approximation ratio induction}
    \label{fig:gma0}
\end{figure*}

\begin{lemma}\label{lemma:tgc2}
    Suppose function $\mathbf{M_z}$ is any (integral) matching of the constructed weighted tripartite graph $\mathcal{H} = (\mathcal{V}_1 \cup \mathcal{V}_2 \cup \mathcal{V}_3, \mathcal{E})$. We can obtain the following conclusions:
    \begin{align*}
        \sum_{i=1}^I\sum_{r=1}^{n_j} \sum_{w_{ks} \in \mathcal{V}_3} \scaleto{M_\mathbf{z}(v_i, w_{jr}, w_{ks})b(v_i, w_{jr}, w_{ks}) \le b_j, \forall j \in \mathcal{J};}{10pt}\\
        \sum_{i=1}^I\sum_{w_{jr} \in \mathcal{V}_2}\sum_{s=1}^{n_k} \scaleto{M_\mathbf{z}(v_i, w_{jr}, w_{ks})c(v_i, w_{jr}, w_{ks}) \le c_k, \forall k \in \mathcal{K}.}{10pt}
    \end{align*}
\end{lemma}
\begin{proof}
    Based on a matching $\mathbf{M_{z}}$ for the tripartite graph $\mathcal{H}$, we can easily construct a matching $\mathbf{M_{x}}$ for the bipartite graph $\mathcal{B}_{\mathbf{\tilde{x}}}$ by letting 
    \[M_\mathbf{x}(v_i, w_{jr}) = \sum_{w_{ks} \in \mathcal{V}_3} M_\mathbf{z}(v_i, w_{jr}, w_{ks}).\]
    Similarly, we can construct a matching $\mathbf{M_{y}}$ for the bipartite graph $\mathcal{B}_{\mathbf{\tilde{y}}}$ by letting 
    \[M_\mathbf{y}(v_i, w_{ks}) = \sum_{w_{jr} \in \mathcal{V}_2} M_\mathbf{z}(v_i, w_{jr}, w_{ks}).\]
    Since $b(v_i, w_{jr}, w_{ks}) = b(v_i, w_{jr})$ and $c(v_i, w_{jr}, w_{ks}) = c(v_i, w_{ks})$ according to line $13$ of $\mathtt{WTGConstruct}$, this lemma can be proved following the result of Lemma \ref{lemma:bgc02}.
\end{proof}

\textit{By combining Lemma~\ref{lemma:tgc1} and Lemma~\ref{lemma:tgc2}, we can conclude that any matching of $\mathcal{H}$ satisfies the deadline, offloading power, and resource constraints in $\mathsf{DTRP}$.} As each task $i$ has only one corresponding task node $v_i$, and every edge $(v_i, w_{jr}, w_{ks})$ of $\mathcal{H}$ corresponds to a task-to-AP-server mapping $(i,j,k)$, a matching of $\mathcal{H}$ can be converted into a feasible solution for $\mathsf{DTRP}$. We first define a relaxed maximum weighted $3$-dimensional matching problem, which aims to identify a fractional matching $F_{\mathbf{z}}$ for $\mathcal{H}$ with maximum total weight. Let $e$ represent edge $(v_i, w_{jr}, w_{ks})$, and $\mathcal{V} = \mathcal{V}_1 \cup \mathcal{V}_2 \cup \mathcal{V}_3$. We denote this LP problem as $\mathsf{3DM}$ and formulate it as follows.
\begin{equation}\label{eq:3dm}
    (\mathsf{3DM}) \ \ \max \textstyle \sum\nolimits_{e \in \mathcal{E}}F_{\mathbf{z}}(e)u(e)
\end{equation}
subject to: 
\addtocounter{equation}{-1}
\begin{subequations}
	\allowdisplaybreaks
	\begin{align}
        \textstyle \sum\nolimits_{e \in \{e' : e' \in \mathcal{E}, v \in e'\}} F_{\mathbf{z}}(e) \le 1, \forall v \in \mathcal{V} \label{eq:3dma}
        \\
        F_{\mathbf{z}}(e) \ge 0, \forall e \in \mathcal{E} \label{eq:3dmb}
	\end{align}
\end{subequations}
Eq.~\eqref{eq:3dma} ensures that the total fractions ($F_{\mathbf{z}}(e)$) of edges incident on node $v$ do not exceed $1$ for any $v \in \mathcal{V}$. We can derive the following lemma based on the formulation of $\mathsf{3DM}$.
\begin{lemma}\label{lemma:tgc3}
    $OPT_\mathsf{RDP} \le OPT_\mathsf{3DM}$
\end{lemma}

\begin{proof}
    In Algorithm~\ref{alg:tgc}, we construct edges for all $\tilde{z}_{ijmkn} > 0$. Thus, for an optimal solution $\tilde{\mathbf{z}}$ of $\mathsf{RDP}$, we can construct a feasible solution $\mathcal{M}_{\mathbf{z}}$ of $\mathsf{3DM}$ that satisfies
    \[ {\textstyle \sum\nolimits_{r=1}^{n_j}\sum\nolimits_{s=1}^{n_k}\mathcal{M}_{\mathbf{z}}(v_i, w_{jr}, w_{ks}) = \sum\nolimits_{m=0}^{\pi_j}\sum\nolimits_{n=0}^{\lambda_k} \tilde{z}_{ijmkn}}.\]
    Since the objective value corresponding to this constructed $\mathcal{M}_{\mathbf{z}}$ is no greater than $OPT_\mathsf{3DM}$, Lemma~\ref{lemma:tgc3} follows.
\end{proof}

Here, we employ an algorithm introduced by Chan and Lau, namely the $\boldsymbol{k}$-Dimensional Matching Algorithm ($\mathtt{kDMA}$) \cite{chan2012linear}, to convert the optimal fractional solution for $\mathsf{3DM}$ into a matching of $\mathcal{H}$. In $\mathtt{kDMA}$, Chan and Lau first solve $\mathsf{3DM}$, and sort all edges $e$ with positive $F_\mathbf{z}(e)$ based on a partial ordering. Then, a recursive function is applied to the sorted list. In each recursive call, one edge $e$ is considered, and the marginal utility of remaining edges is updated based on the selection decision of $e$. This recursive function eventually returns a (integral) matching $\mathbf{M}_{\mathbf{z}}$ for $\mathcal{H}$. (The details of $\mathtt{kDMA}$ are presented in Appendix \ref{app:3kdm}.) 

\textbf{Time Complexity.} In $\mathtt{WTGConstruct}$, there are at most $IJUKV$ positive $\tilde{z}_{ijkmn}$, which defines at most $4IJUKV$ edges in the weighted tripartite graph $\mathcal{H}$. Therefore, $\mathsf{3DM}$ has at most $4IJUKV$ edges, solving which takes $\mathcal{O}((IJUKV)^3)$ time \cite{vaidya1987an}. The edge sorting operation takes $\mathcal{O}((IJUKV)^2)$. In the recursive function, there are at most $4IJUKV$ recursive layers. In each recursive layer, the marginal utilities of at most $4IJUKV$ remaining edges are updated. Therefore, the time complexity of algorithm $\mathtt{kDMA}$ is $\mathcal{O}((IJUKV)^3)$. Based on their findings, we present the following proposition.

\begin{proposition}\label{prop:trigc1}
    ({Theorem 2.6}, \cite{chan2012linear}) $\mathtt{kDMA}$ can obtain a matching $\mathbf{M}_{\mathbf{z}}$ of $\mathcal{H}$ from the optimal solution of $\mathsf{3DM}$, satisfying the condition 
    \[\textstyle{OBJ_{\mathtt{GMA}} = \sum_{e \in \mathcal{E}}M_{\mathbf{z}}(e)u(e) \ge \frac{1}{2}OPT_{\mathsf{3DM}}}.\]
\end{proposition}

Next, we prove the theoretical guarantee of $\mathtt{GMA}$. For ease of understanding, we also provide an algorithm flow of $\mathtt{GMA}$ and its approximation ratio induction overview in Fig. \ref{fig:gma0}.

\begin{theorem}\label{theorem:gaa}
    Let $\epsilon$ denote the resource discretization loss, where $\epsilon > 0$ and $\varphi = 1 + \frac{\epsilon}{2}$ ($\varphi$ is the discretization step defined in Subsection~\ref{subsec:ld}). $\mathtt{GMA}$ yields a feasible task offloading and resource allocation solution $\{\mathbf{x}, \mathbf{y}, \mathbf{b}, \mathbf{c}, \mathbf{p}\}$ for $\mathsf{DTRP}$ with an objective value $OBJ_{\mathtt{GMA}} \ge \frac{1-\alpha}{2+\epsilon}OPT_{\mathsf{DTRP}}$.
\end{theorem}
\begin{proof}
    For each $\mathcal{M}_{\mathbf{z}}(v_i, w_{jr}, w_{ks}) = 1$, $\mathtt{GMA}$ sets $x_{ij} = 1$, $y_{ik} = 1$, $b_{ij} = b(v_i, w_{jr}, w_{ks})$, and $c_{ik} = c(v_i, w_{jr}, w_{ks})$. As each task only has one corresponding task node in graph $\mathcal{H}$, at most one edge that contains node $v_i$ can be selected for each task $i$ in a matching of $\mathcal{H}$. Thus,
    $OBJ_{\mathtt{GMA}} = \sum_{e \in \mathcal{E}}M_{\mathbf{z}}(e)u(e)$, and the resulting solution $\{\mathbf{x}, \mathbf{y}, \mathbf{b}, \mathbf{c}, \mathbf{p}\}$ satisfies constraints~\eqref{eq:trpb}$\sim$\eqref{eq:trpd} of $\mathsf{DTRP}$. Based on Lemma~\ref{lemma:tgc1} and Lemma~\ref{lemma:tgc2}, the resulting solution also satisfies the deadline, offloading power and resource constraints of  $\mathsf{DTRP}$. Therefore, the resulting solution is a feasible solution for $\mathsf{DTRP}$.
    
    According to Proposition~\ref{prop:trigc1}, $OBJ_{\mathtt{GMA}} \ge \frac{1}{2}OPT_\mathsf{3DM}$. Besides, based on Lemma~\ref{lemma:tgc3} ($OPT_\mathsf{3DM} \ge OPT_\mathsf{RDP}$) and Lemma~\ref{lemma:ld1} ($OPT_\mathsf{RDP} \ge \frac{1-\alpha}{\varphi}OPT_\mathsf{DTRP}$), we can conclude 
    \[{\textstyle {OBJ_\mathtt{GMA} \ge \frac{1}{2}OPT_\mathsf{3DM} \ge \frac{1}{2}OPT_\mathsf{RDP} \ge \frac{1-\alpha}{2\varphi}OPT_\mathsf{DTRP}}.}\]
    Substituting $\varphi = 1 + \frac{\epsilon}{2}$, we get $OBJ_\mathtt{GMA} \ge \frac{1-\alpha}{2+\epsilon}OPT_\mathsf{DTRP}$. 
    In Subsection~\ref{subsec:ld}, we set $\pi_j = \left\lceil \log_{\varphi}(\alpha b_j) \right\rceil$. To ensure $\pi_j$ is polynomial in the size of input $ b_j$, $\varphi$ should be strictly greater than $1$. As a result, $\epsilon > 0$.
\end{proof}

\textbf{$\mathtt{GMA}$ Time Complexity Analysis}: 
\begin{itemize}
    \item Line $1$: The LP problem $\mathsf{RDP}$ has at most $IJUKV$ variables, solving which takes $\mathcal{O}((IJUKV)^3)$ time \cite{vaidya1987an}.
    \item Line 2: Using $\mathtt{WTGConstruct}$ to construct the tripartite graph takes $\mathcal{O}(IJUKV\log(IJUKV))$ time.
    \item \textit{Line 3}: The time complexity of $\mathtt{kDMA}$ is $\mathcal{O}((IJUKV)^3)$. Therefore, the time complexity for obtaining a matching $\mathcal{H}$ is $\mathcal{O}((IJUKV)^3)$.
    \item \textit{Line $4$--$7$}: The time complexity is $\mathcal{O}(I)$ since at most $I$ tasks can be offloaded.
\end{itemize}
As a result, the time complexity of $\mathtt{GMA}$ is $\mathcal{O}((IJUKV)^3)$. In Subsection~\ref{subsec:ld}, we introduce a logarithmic function to discretize the resource allocation. This log-based discretization ensures that $U$ and $V$ grow polynomially with respect to the input sizes of $b_j$ and $c_k$. Hence, the overall time complexity of $\mathsf{GMA}$ becomes polynomial in the input size of $\mathsf{DTRP}$.

\textbf{Discussion.} In $\mathtt{GMA}$, we incorporate the tripartite graph matching algorithm proposed by Chan and Lau~\cite{chan2012linear}. However, their method alone does not determine task-specific resource allocations. To address this, $\mathtt{GMA}$ introduces a novel combination of resource discretization and tripartite graph construction, enabling joint optimization of task mapping (to both APs and servers), resource allocation (for both offloading and processing), and offloading power control. This constitutes the first approximation algorithm for $\mathsf{DTRP}$ with polynomial-time complexity. In real-world cloud infrastructures, resource allocation bounds are typically tight. For instance, Alibaba datacenters allow servers with over $96$ logical CPU cores, yet cap per-task allocations at $16$ cores~\cite{weng2022mlaas}. Similarly, Google Cloud Run~\cite{google_cpu_limit}, Azure Functions~\cite{azure_cpu_limit}, and AWS Lambda~\cite{aws_cpu_limit} limit allocations to $8$, $4$, and $6$ cores, respectively. These configurations imply that the resource allocation bound $\alpha$ is generally no greater than $\frac{1}{6}$ in practice, making the approximation ratio of $\mathtt{GMA}$ close to $\frac{1}{2}$ (since $\epsilon$ is a small positive constant). Finally, we note that $\mathsf{DTRP}$ is significantly more complex than GAP, for which the best-known and tight approximation ratio is $\frac{1}{2}$~\cite{shmoys1993approximation}. Therefore, achieving a provable bound exceeding $\frac{1}{2}$ for $\mathsf{DTRP}$ is unlikely. These insights suggest that $\mathtt{GMA}$ offers a practical deterministic bound approaching its maximum likely approximation ratio of $\frac{1}{2}$.

\section{Experimental Evaluation}\label{sec:experiment}
\begin{figure*}[htb!]
    \centering
    \includegraphics[page=1, width=1\textwidth]{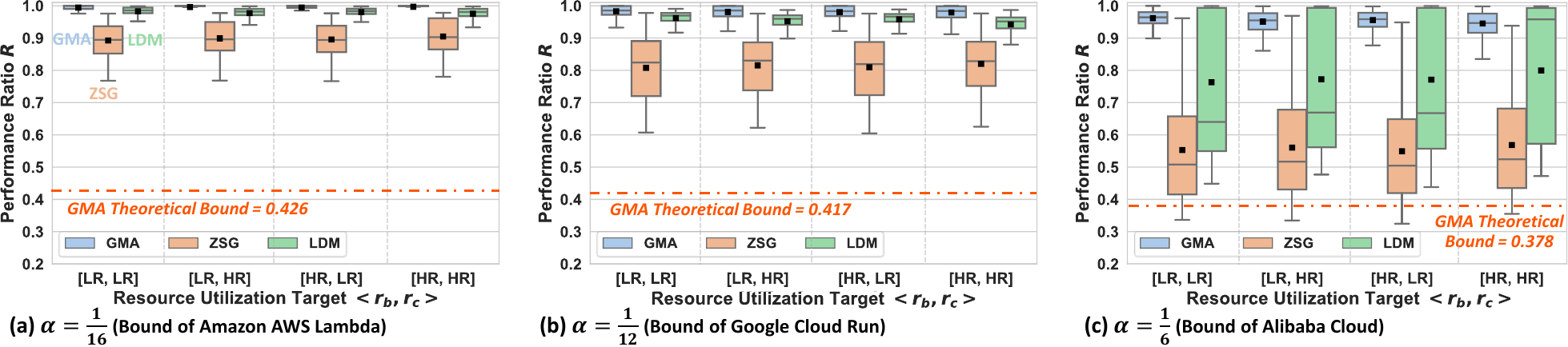}
    \caption{Performance Ratio $(R)$ by different algorithms (GMA, ZSG, LDM) under varying resource allocation bounds $\alpha$}
    \label{fig:exp-pg}
\end{figure*}

In this section, we evaluate the performance of $\mathtt{GMA}$ through numerical simulations. To ensure a comprehensive assessment, we compare $\mathtt{GMA}$ with two existing heuristic algorithms based on the achieved performance ratio $R$, defined as
\[R = \frac{\text{total saved energy by the algorithm}}{\text{total saved energy by an optimal policy}}.\]
\textbf{Optimal Policy}: Since obtaining the exact solution to $\mathsf{DTRP}$ is intractable, we formulate a new LP problem derived from $\mathsf{RDP}$ by replacing $(1-\alpha)$ in Eqs.~\eqref{eq:rdpb} and \eqref{eq:rdpc} with the discretization factor $\varphi$. It can be shown (similar to the proof of Lemma~\ref{lemma:ld1}) that the optimal solution of this LP problem is at least as large as $OPT_\mathsf{DTRP}$. Therefore, we adopt this solution as the optimal policy for computing $R$. Since the denominator is no less than $OPT_\mathsf{DTRP}$, the computed value of $R$ underestimates the true performance ratio based on the actual optimal saved energy $OPT_\mathsf{DTRP}$. As a result, the values reported in Figs.~\ref{fig:exp-pg} and \ref{fig:exp4} serve as lower bounds on the actual performance that the algorithms can achieve.

\subsection{MEC Architecture and Taskset Generation}
We sample system-related parameters and tasksets from ranges considered in existing studies and practical systems. 
The numbers of APs and servers are fixed at $12$ and $15$, respectively. $12$ of the $15$ servers are co-located with APs.
We sample $c_k$ of each server $k \in \mathcal{K}$ within  $20 \sim 30$ Giga cycles/s \cite{xiang2021dataset}, and choose $b_j$ of each AP $j \in \mathcal{J}$ from $\{80, 120\}$ MHz (802.11n Wi-Fi Protocol) \cite{intel_wifi_protocal}. The backhaul network delay $\delta_{jk}$ is sampled from $3 \sim 30$ ms \cite{alameddine2019dynamic}. We set the wireless network channel gain $G_{ij}$ as $-50$ dB \cite{zhou2018computation}, noise power $\sigma^2$ as $8e^{-8}$, and maximum offloading power $p_{max}$ as $0.1$ W \cite{li2022dynamic}. 

We generate tasksets with varying resource utilization targets $\langle r_b,r_c\rangle$, and taskset size $I$. The resource utilization target $r_b$ ($r_c$) of a taskset is the ratio of the total targeted communication (computation) resource demand of all tasks in a taskset to the total communication (computation) resource of the system~\cite{gao2022deadline}. We consider two ranges, $LR = [0.7, 1]$ and $HR = [1.2, 1.5]$, and sample $\langle r_b,r_c\rangle$ from four different range combinations: $\langle LR, LR\rangle$, $\langle LR, HR\rangle$, $\langle HR, LR\rangle$, and $\langle HR, HR\rangle$. For each range combination, we sample $30$ different $\langle r_b,r_c\rangle$ values. For each sampled $\langle r_b,r_c\rangle$, we sample $30$ different $I$ from $[50, 200]$.

Given the values for $\langle r_b,r_c\rangle$ and $I$, we generate a single taskset as follows. 
We randomly sample $s_i$ in $[100, 200]$ Kb that matches the size of a typical image and set $\eta_i = 150$ for each task $i$ \cite{xu2020energy}. Next, we sample the local computation resource capacity $f_i$ within $[1,2]$ Giga cycles/s~\cite{li2022dynamic}. Given $r_b$ and $r_c$, let $R_b = r_b \sum_{j \in \mathcal{J}b_j}$ and $R_c = r_c \sum_{k \in \mathcal{K}}c_k$ be the targeted bandwidth and computation resource demand of the taskset, respectively. Then, we use Stafford's Randfixedsum Algorithm~\cite{emberson2010techniques} to distribute $R_b$ and $R_c$ to each individual task in the taskset in a uniformly random and unbiased manner. Let $R_{b}^i$ and $R_{k}^i$ be the assigned targeted bandwidth and computation resource demand of task $i$. We set $d_i$ with $d_i = \frac{s_i}{r_{ij}} + \delta_i + \frac{s_i\eta_i}{R_{k}^i}$, where $\delta_i \sim \mathcal{N}(8,3)$ ms covers half of the sample range of $\delta_{jk}$, and $r_{ij}$ is computed based on Eq. \eqref{eq:model3} with $p_{ij}^o = p_{max}$ and $b_{ij}=R_{b}^i$. For each taskset, we choose $|\mathcal{J}_i|$ in $\{2,3\}$ and randomly assign tasks to APs, while ensuring that the number of tasks that can be offloaded to each AP is drawn from a normal distribution. This, combined with Stafford's Randfixedsum Algorithm for tasks' targeted resource demand assignment, ensures we generate various distributions of workload demand and their assignment to APs, in an unbiased manner.

\textit{Baseline Algorithms.} We employ two heuristic algorithms proposed by Gao \emph{et al.}~\cite{gao2022deadline}, namely \texttt{ZSG} and \texttt{LDM}, as the baseline algorithms. Their study proposed a similar MEC architecture, and jointly considered task mapping (to both APs and servers) and resource allocations (for both offloading and processing) for deadline-constrained tasks, with the aim of maximizing user-defined profit.
\begin{enumerate}
    \item For each task mapping, \texttt{ZSG} estimates the time for task offloading and task processing based on the task's data size and required compute cycles, which are then used to compute the resource allocations. Then, \texttt{ZSG} greedily selects the mapping and resource allocation with the highest energy-to-resource allocation ratio whenever the system possesses sufficient resources.
    \item \texttt{LDM} reformulates $\mathsf{DTRP}$ into an Integer Linear Programming (ILP) problem by discretizing the system's resource allocations using \textit{equal-sized intervals}, where the interval sizes of $1$ MHz and $50$ Mega cycles/s are used for bandwidth and computational resources, respectively. Then, an ILP solver is employed to solve the ILP problem.
\end{enumerate}

We set $\epsilon$ as $0.2$ for $\mathtt{GMA}$, and allocate the same runtime for \texttt{LDM} as $\mathtt{GMA}$. For each taskset, we run $\mathtt{GMA}$, \texttt{ZSG}, and \texttt{LDM} with varying $\alpha$ in $\{\frac{1}{16}, \frac{1}{12}, \frac{1}{6}\}$ (bounds obtained from Amazon AWS Lambda, Google Cloud Run, and Alibaba Cloud). Thus, each algorithm runs $10800$ simulations. Experiments were conducted on a desktop PC with an Intel(R) Xeon(R) E7-8880V4 2.20GHz CPU and 32GB of RAM.

\subsection{Performance Evaluation}

\begin{figure}[t]
    \centering
    \includegraphics[page=5, width=0.9\columnwidth]{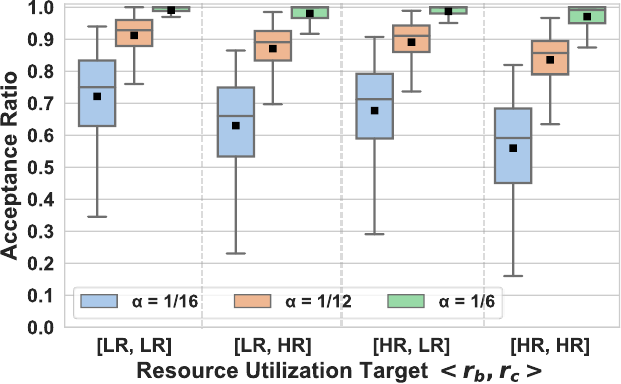}
     \caption{Achieved Acceptance Ratio by $\mathtt{GMA}$}
     \label{fig:exp4}
\end{figure}


We evaluate the performance of different algorithms by comparing the achieved performance ratio $R$ under various $\alpha$ values (Fig.~\ref{fig:exp-pg}). (Note that the line in the middle of the box refers to the median value, and the black square dot refers to the average value.) $\mathtt{GMA}$ achieves an average performance ratio of $99.5\%$ for $\alpha = \frac{1}{16}$, $98.1\%$ for $\alpha = \frac{1}{12}$, and $95.3\%$ for $\alpha = \frac{1}{6}$. Considering $\epsilon = 0.2$, $\mathtt{GMA}$ provides a theoretical bound of $0.426$ for $\alpha = \frac{1}{16}$, $0.417$ for $\alpha=\frac{1}{12}$, and $0.378$ for $\alpha = \frac{1}{6}$. As a result, $\mathtt{GMA}$'s practical performance surpasses its theoretical bounds by an average of $56.93\%$, and shows its stability under varying resource allocation bounds and resource usage levels. The results also show that $\mathtt{GMA}$ effectively bridges the gap between its theoretical bound and the optimal solution, showing the practical efficiency of $\mathtt{GMA}$.

$\mathtt{GMA}$ exhibits an average performance ratio that is $22.04\%$  and $7.36\%$ higher than \texttt{ZSG} and \texttt{LDM}, respectively. Notably, unlike $\mathtt{GMA}$, both \texttt{ZSG} and \texttt{LDM} do not have any theoretical guarantees. Compared to \texttt{ZSG}, $\mathtt{GMA}$ performs significantly better and has far less variability. Besides, the performance of \texttt{LDM} is comparable with $\mathtt{GMA}$ when $\alpha$ equals $ \frac{1}{16}$ and $\frac{1}{12}$, but degrades and becomes highly variable when $\alpha = \frac{1}{6}$. Thus, in conclusion, $\mathtt{GMA}$ provides the first algorithm for solving $\mathsf{DTRP}$ with a theoretical approximation bound and excellent performance with low variation for practical systems.

We also evaluate the achieved acceptance ratio by $\mathtt{GMA}$ under various $\alpha$ values (Fig.~\ref{fig:exp4}). The acceptance ratio is the ratio of the number of tasks being offloaded to the number of tasks in the taskset. $\mathtt{GMA}$ achieves an average acceptance ratio of $64.7\%$ for $\alpha=\frac{1}{16}$, $87.7\%$ for $\alpha=\frac{1}{12}$, and $98.2\%$ for $\alpha=\frac{1}{6}$. As illustrated in Fig.~\ref{fig:exp4}, the acceptance ratio exhibits a notable sensitivity to the choice of $\alpha$, where a higher $\alpha$ corresponds to an elevated average acceptance ratio. This correlation arises because a larger $\alpha$ increases the resources that can be allocated to each task, thereby enhancing their ability to meet deadlines and resulting in an elevated acceptance ratio. Furthermore, our observations indicate that an increasing resource utilization target $\langle r_b, r_c \rangle$ is associated with a marginal decrease in the acceptance ratio. This phenomenon arises due to the inherent complexity of provisioning tasksets with higher resource utilization targets, making them comparatively more challenging to accommodate.

\section{Conclusion}\label{sec:conclusion}
This paper investigated the deadline-constrained task offloading and resource allocation problem in MEC with both communication and computation resource contentions. 
In this general system, we jointly optimized task mapping to both APs and servers, resource allocation for offloading and processing, and dynamic offloading power control. To address this problem, we proposed the Graph-Matching-based Approximation Algorithm ($\mathtt{GMA}$), the first polynomial-time approximation algorithm of its kind. $\mathtt{GMA}$ achieves a provable approximation ratio of $\frac{1-\alpha}{2+\epsilon}$, where $\alpha$ is the resource allocation bound and $\epsilon$ is a small positive constant. Experimental results demonstrated that $\mathtt{GMA}$ consistently outperforms existing baseline algorithms in both effectiveness and stability.

For future work, we plan to investigate the resource augmentation problem in MEC. Specifically, given that a feasible solution exists that admits all tasks, the goal is to find a solution that minimizes overall system resource usage while ensuring all tasks are still successfully admitted.

\appendices
\section{Fractional Matching and Matching}\label{app:matching}
\begin{figure}[h]
    \centering
    \includegraphics[width=1\linewidth, page=1]{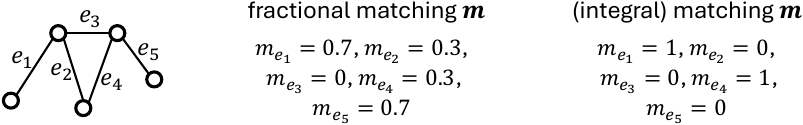}
    \caption{An example of a fractional matching and a (integral) matching of a graph with $5$ nodes and $5$ edges.}
    \label{fig:matching}
\end{figure}

Consider a graph $\mathcal{G} = (\mathcal{V}, \mathcal{E})$, where $\mathcal{V}$ is the node set and $\mathcal{E}$ is the edge set. A fractional matching \cite{aharoni1990possible} of $\mathcal{G}$ is a function $\mathbf{m}$ that assigns each edge $e \in \mathcal{E}$ with a fraction $m_e$ in the range of $[0, 1]$, such that for every node $v \in \mathcal{V}$, the total fractions of edges incident on $v$ is at most $1$, i.e., $\sum_{e: v \in e, e \in \mathcal{E}}m_e \le 1, \forall v \in \mathcal{V}$.
If $m_e \in \{0,1\}$ for each edge in $\mathcal{E}$, $\mathbf{m}$ is a (integral) matching \cite{lovasz2009matching} of $\mathcal{G}$. An example is provided in Fig. \ref{fig:matching}.

\section{Bipartite Graph}\label{app:bipartite}
A bipartite graph \cite{lovasz2009matching} $\mathcal{B} = (\mathcal{V} ,\mathcal{W}, \mathcal{E})$ is a graph whose nodes can be divided into two disjoint and independent sets $\mathcal{V}$ and $\mathcal{W}$, and every edge in the edge set $\mathcal{E}$ connects a node in $\mathcal{V}$ to a node in $\mathcal{W}$. Besides, no edge connects two nodes in the same set. An example of a bipartite graph is provided on the left of Fig. \ref{fig:bipartite}, where each edge comprises two nodes, one from each partitioned node set.

\section{Weighted Tripartite Graph}\label{app:tripartite}
\begin{figure}[h]
    \centering
    \includegraphics[width=0.6\linewidth, page=2]{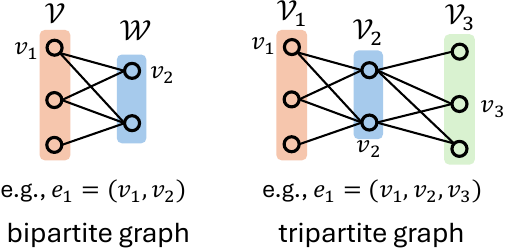}
    \caption{An example of a bipartite graph (left) and a tripartite graph (right).}
    \label{fig:bipartite}
\end{figure}

A tripartite graph \cite{chan2012linear} $\mathcal{H} = (\mathcal{V}, \mathcal{E})$ is a hypergraph whose node set $\mathcal{V}$ can be partitioned into three disjoint and independent sets, $\mathcal{V}_1, \mathcal{V}_2$ and $ \mathcal{V}_3$. Each edge $e \in \mathcal{E}$ is a subset of $\mathcal{V}$, which contains exactly three nodes and intersects each partitioned set ($\mathcal{V}_1, \mathcal{V}_2$ or $ \mathcal{V}_3$) in exactly one node. A \textit{weighted tripartite graph} is a tripartite graph where each edge $e \in \mathcal{E}$ is associated with a real-valued weight $u_e$. An example of a tripartite graph is provided on the right of Fig. \ref{fig:bipartite}. 

\section{k-Dimensional Matching Algorithm}\label{app:3kdm}
\begin{algorithm}[h]
    \caption{$\mathtt{kDMA}$ \cite{chan2012linear}}\label{alg:3kdm}
    \SetAlgoNoEnd
    \SetKwInOut{Input}{input}
    \SetKwInOut{Output}{output}
    \SetKwFor{ForEach}{for each}{do}{endfch}
    \SetKw{KwAnd}{and}
    \SetKwFunction{FLOCAL}{\textbf{LocalRatio}}
    \SetKwProg{Fn}{Function}{:}{}
    \SetNoFillComment

    Find an optimal basic solution $\mathbf{F}_\mathbf{z}$ to $\mathsf{3DM}$. Remove every hyperedge $e$ from $\mathcal{E}$ with $F_\mathbf{z}(e)=0$. Initialize $\mathcal{Q} \leftarrow \emptyset$\;
    \For{$s \leftarrow 1$ to $|\mathcal{E}|$}{
    \tcc{\textcolor{darkgray}{\footnotesize{Let $\mathcal{N}(e)$ denote the set of hyperedges that intersect $e$, including $e$ itself}}}
    Find a hyperedge $e$ with $F_\mathbf{z}(\mathcal{N}(e)) \le 2$\;
    Add $e$ to the end of $\mathcal{Q}$, and remove it from $\mathcal{E}$\;
    Remove $F_\mathbf{z}(e)$ from $\mathbf{F}_\mathbf{z}$\;
    }
    $\mathbf{M_z}\leftarrow$ \FLOCAL{$\mathcal{Q}, \mathbf{u}$}\;
    \textbf{return} $\mathbf{M_z}$\;
    
    \BlankLine
    \tcc{\textcolor{darkgray}{\small{a recursive subroutine}}}
    \Fn{\FLOCAL{$\mathcal{Q}, \mathbf{u}$}}{
        Remove from $\mathcal{Q}$ all hyperedges with non-positive weights\;
        \lIf{$\mathcal{Q}=\emptyset$}{\textbf{return} $\emptyset$}
        Choose the leftmost hyperedge $e$ from updated $\mathcal{Q}$. Decompose the weight vector $\mathbf{u}=\mathbf{u_1} + \mathbf{u_2}$ where
        \begin{equation*}
            u_1(e')=
            \begin{cases}
                u(e) & \text{if } e' \in \mathcal{N}(e), \\
                0 & \text{otherwise}.
            \end{cases}
        \end{equation*}
        $\mathcal{S}' \leftarrow$ \FLOCAL{$\mathcal{Q}, \mathbf{u_2}$}\;
        \scalebox{1}{\lIf{$\mathcal{S}' \cup \{e\}$ is a matching}{\textbf{return} $\mathcal{S} = \mathcal{S}' \cup \{e\}$}}
        \lElse{\textbf{return} $\mathcal{S} = \mathcal{S}'$}
    }
\end{algorithm}

\bibliographystyle{IEEEtran}
\bibliography{IEEEabrv,references}

\end{document}